\documentclass[acmsmall]{acmart}

\usepackage{balance}
\usepackage{hyperref}
\usepackage{xcolor}
\usepackage{multirow}
\usepackage{graphicx}
\usepackage{amsmath}
\usepackage{fontawesome5}
\usepackage{makecell}
\usepackage{pdflscape}
\usepackage{adjustbox}
\usepackage{afterpage}
\usepackage{array}
\usepackage{siunitx}
\usepackage{longtable}
\usepackage{tikz}

\usepackage[absolute,showboxes]{textpos}

\makeatletter
\newcommand{\paragraphN}[1]{\vspace*{0.03in}\noindent{\bf #1}\hspace{0.25ex \@plus1ex \@minus.2ex}}
\makeatother

\let\orgautoref\autoref
\renewcommand{\autoref}
{\def\sectionautorefname{Section}\def\subsectionautorefname{Section}\def\subsubsectionautorefname{Section}\def\equationautorefname{Eq.}\orgautoref}

\usepackage{pifont}

\usepackage{xspace}
\newcommand{\cf}{\textit{cf.}~}
\newcommand{\etal}{\textit{et al.}~}
\newcommand{\eg}{\textit{e.g.,}~}
\newcommand{\ie}{\textit{i.e.,}~}

\newcommand{\vs}{\textit{vs.}~}
\newcommand{\one}{({\em i})\xspace}
\newcommand{\two}{({\em ii})\xspace}
\newcommand{\three}{({\em iii})\xspace}

\makeatletter
\renewcommand{\paragraph}[1]{\vspace*{0.03in}\noindent{\bf #1.}\hspace{0.25ex \@plus1ex \@minus.2ex}}
\makeatother

\makeatletter
\newcommand{\paragraphc}[1]{\vspace*{0.03in}\noindent{\bf #1}\hspace{0.25ex \@plus1ex \@minus.2ex}}
\makeatother

\newcommand\Tstrut{\rule{0pt}{2.6ex}}       \newcommand\Bstrut{\rule[-0.9ex]{0pt}{0pt}} \newcommand{\TBstrut}{\Tstrut\Bstrut}

\newcolumntype{?}{!{\vrule width 1pt}}
\newcolumntype{x}[1]{>{\centering\arraybackslash\hspace{0pt}}p{#1}}

\usepackage{xcolor}
\definecolor{colorDesyPetrol}{HTML}{006987}
\definecolor{darkGray}{HTML}{333333}

\newcommand*\desyPetrolCircle[1]{\protect\tikz[baseline=(char.base)]{
    \protect\node[shape=circle, draw=white, anchor=south, minimum size=1pt, inner sep=1.2pt, fill=darkGray, font=\small] (char) {\bfseries{\textsf{#1}}};}}

\newcommand*\desyPetrolCircleEmpty[1]{\protect\tikz[baseline=(char.base)]{
    \protect\node[shape=circle, draw=darkGray, anchor=south, minimum size=1pt, line width=0.8pt, inner sep=0.8pt, fill=white, font=\small] (char) {\bfseries{\textsf{#1}}};}}

\begin{document}

\fancypagestyle{firstpagestyle}{%
    \fancyfoot[RO,LE]{\footnotesize ACM Computing Surveys}%
}
\fancyfoot[RO,LE]{\footnotesize ACM Computing Surveys}%

\setlength{\TPHorizModule}{\paperwidth}
\setlength{\TPVertModule}{\paperheight}
\TPMargin{5pt}
\begin{textblock}{0.8}(0.1,0.02)
     \noindent
     \footnotesize
     If you cite this paper, please use the ACM CSUR reference:
     Leandro Lanzieri, Gianluca Martino, Goerschwin Fey, Holger Schlarb, Thomas C. Schmidt, and Matthias W{\"a}hlisch. 2024.
     A Review of Techniques for Ageing Detection and Monitoring on Embedded Systems.
     In \emph{ACM Comput. Surv.} https://doi.org/10.1145/3695247.
\end{textblock}

\newcommand{\paperTitle}{A Review of Techniques for Ageing Detection and Monitoring on Embedded Systems}
\title{\paperTitle}

\author[L. Lanzieri]{Leandro Lanzieri}
\email{leandro.lanzieri@tuhh.de}
\orcid{https://orcid.org/0000-0002-9804-7258}
\affiliation{\department{Institute of Embedded Systems}
\institution{Hamburg University of Technology}
\city{Hamburg}
\country{Germany}
}
\additionalaffiliation{
  \institution{Deutsches Elektronen-Synchrotron DESY and HAW Hamburg, Germany}
  \city{Hamburg}
  \country{Germany}
}

\author[G. Martino]{Gianluca Martino}
\email{gianluca.martino@tuhh.de}
\orcid{https://orcid.org/0000-0002-7838-3844}
\affiliation{\department{Institute of Embedded Systems}
  \institution{Hamburg University of Technology}
  \city{Hamburg}
  \country{Germany}
}

\author[G. Fey]{Goerschwin Fey}
\email{goerschwin.fey@tuhh.de}
\orcid{https://orcid.org/0000-0001-6433-6265}
\affiliation{\department{Institute of Embedded Systems}
  \institution{Hamburg University of Technology}
  \city{Hamburg}
  \country{Germany}
}

\author[H. Schlarb]{Holger Schlarb}
\email{holger.schlarb@desy.de}
\orcid{https://orcid.org/0000-0003-4115-5183}
\affiliation{\institution{Deutsches Elektronen-Synchrotron DESY}
  \city{Hamburg}
  \country{Germany}
}

\author[T.C. Schmidt]{Thomas C. Schmidt}
\email{t.schmidt@haw-hamburg.de}
\orcid{https://orcid.org/0000-0002-0956-7885}
\affiliation{\department{Department Informatik}
  \institution{HAW Hamburg}
  \city{Hamburg}
  \country{Germany}
}

\author[M. W{\"a}hlisch]{Matthias W{\"a}hlisch}
\email{m.waehlisch@tu-dresden.de}
\orcid{https://orcid.org/0000-0002-3825-2807}
\affiliation{\department{Faculty of Computer Science}
  \institution{TU Dresden and Barkhausen Institut}
  \city{Dresden}
  \country{Germany}
}

\begin{abstract}
  Embedded digital devices are progressively deployed in dependable or safety-critical systems.
  These devices undergo significant hardware ageing, particularly in harsh environments. This increases their likelihood of failure.
  It is crucial to understand ageing processes and to detect hardware degradation early for guaranteeing system dependability.
  In this survey, we review the core ageing mechanisms,  identify and categorize general working principles of ageing detection and monitoring techniques for Commercial-Off-The-Shelf (COTS) components that are prevalent in embedded systems: Field Programmable Gate Arrays (FPGAs), microcontrollers, System-on-Chips (SoCs), and their power supplies.
  From our review, we find that online techniques are more widely applied on FPGAs than on other components, and see a rising trend towards machine learning application for analysing hardware ageing.
  Based on the reviewed literature, we identify research opportunities and potential directions of interest in the field.
  With this work, we intend to facilitate future research by systematically presenting all main approaches in a concise way.
\end{abstract}

\setcopyright{acmlicensed}
\acmSubmissionID{CSUR-2023-0178}
\acmJournal{CSUR}
\acmYear{2024}
\acmVolume{}
\acmNumber{}
\acmArticle{}
\acmMonth{9}
\acmDOI{10.1145/3695247}



\maketitle

\begin{acks}
{We acknowledge the support by \grantsponsor{dashh}{DASHH (Data Science in Hamburg - HELMHOLTZ Graduate School for the Structure of Matter)}{}\
	with Grant No.: \grantnum{dashh}{HIDSS-0002}, as well as the support of the \grantsponsor{c-ray4edge}{German Federal Ministry of Education and Research}{C-ray4edge}\  with Grant \grantnum{c-ray4edge}{C-ray4edge}.}
\end{acks}

\begin{figure}[t]
    \centering
    \includegraphics[width=\textwidth]{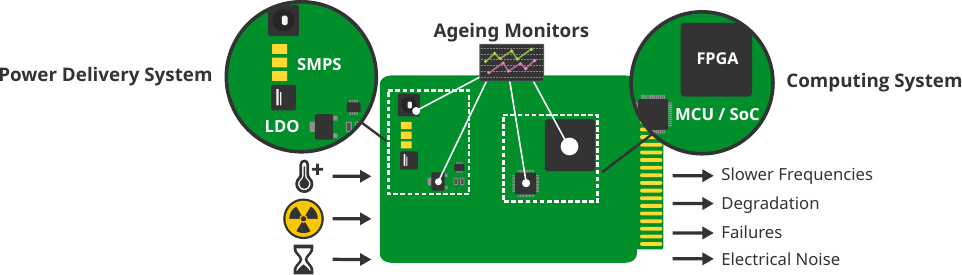}
    \caption{Ageing monitoring is required to ensure the reliability of embedded systems, which are affected by environmental and operational conditions. This survey covers FPGAs, Microcontrollers, and SoCs together with their power supplies as the prevalent system components.}
    \label{fig:concept}
  \end{figure}

\section{Introduction}\label{sec:introduction}
The ubiquity of embedded devices keeps on growing, driven by increasing chip integration levels and computing capacities, as well as by reducing prices and power consumptions.
Applications ranging from consumer electronics and Internet of Things (IoT) gadgets to space missions, nuclear power plants, and particle accelerators involve embedded systems.
Many deployments challenge the lifetime of these devices by making use of unhardened Commercial Off-The-Shelf (COTS) components, by extending mission times over the guaranteed life span, or by operating in harsh conditions.
Nevertheless, properly performant devices are essential for an effective system operation.
Performance is crucial for the timely gathering and processing of data, and essential for accurate decision-making and actuating, for example in industrial environments or in autonomous vehicles.

Microcontrollers and System-on-Chips (SoCs), as well as Field Programmable Gate Arrays (FPGAs) are the most widely deployed digital components in embedded applications. Their prevalent use in exceptionally large multiplicities makes them highly relevant, which motivates the focus of this survey (see \autoref{fig:concept}).

Microcontrollers and SoCs can perform complex tasks at low prices and low power capabilities.
They can even run operating systems with full-featured IPv6 software stacks \cite{bghkl-rosos-18}, which makes them a popular central component in embedded scenarios.
FPGAs offer a high level of flexibility compared to Application Specific Integrated Circuits (ASICs).
Serving even demands from Artificial Intelligence (AI) on the Adaptive Compute Acceleration Platforms (ACAPs), which combine CPU cores, programmable logic, and AI inference acceleration, FPGAs open their way into the edge-computing and high-end edge IoT markets.
All systems require power conditioning and delivery with stable parameters over the duration of their deployment.
Failures in such power delivery modules are usually fatal for the entire system, due to their prominent and critical role.
Nonetheless, it has been reported that the reliability of power systems remains inferior to other modules \cite{trps-zllh-19,epps-ghj-07,psr-p-97}, partially due to their complexity and multitude of comprising components.

Hardware ageing is a severe problem for embedded devices \cite{wtgo-gvgs-18}. Hence, ageing detection plays a key role in dependable or safety-critical applications.
The impact of physical degradation increases for new, highly integrated technologies \cite{rnmm-mp-18,btic-khkr}.
A particular issue for many Integrated Circuits (ICs) in long-term deployments is transistor degradation, namely Bias Temperature Instability (BTI), Hot-Carrier Injection (HCI), and Time-Dependent Dielectric Breakdown (TDDB).
Their predominant consequence is an increment of the transistor threshold voltage, which reduces the maximum switching speed of Complementary Metal-Oxide-Semiconductor (CMOS) logic circuits, which in turn potentially affects critical paths in digital applications and the noise margins on digital gates \cite{enav-kmo-15}.

Embedded systems are composed of a variety of basic components, most of which have been individually analysed \cite{aaec-kcbg-12, lpec-g, rahd-swz, iact-rlnk-17}.
To assess the reliability of embedded systems, a comprehensive approach to detecting failures \cite{wrsw-phamo-21} and monitoring ageing of the entire hardware boards is missing. The present survey closes this gap.

In this paper, we concentrate on techniques that detect degradation processes on COTS components without requiring ageing sensors built in by vendors on the chips. Existing surveys focus on on-chip monitoring solutions \cite{socm-rfes-12} or on reconfiguration and ageing monitoring solutions  \cite{samr-jmmg-20}. This  excludes the wide variety of industrial and commercial deployments that utilize COTS components without dedicated ageing monitoring \cite{mtca-baag-13}.
In contrast, our survey places a special focus on techniques that can be applied to COTS components and that require no special silicon sensors to operate. Khoshavi \etal \cite{ccag-kadk-17} presented and categorized ageing monitoring and mitigation techniques on CMOS devices, focusing  on ring-oscillator sensors.
Kochte \etal \cite{stsa-kw} reviewed literature on self-tests and diagnosis techniques to improve the self-awareness of digital systems.
The authors propose a test classification based on the moment when the self-evaluation takes place in a system. We expand this classification work to address ageing detection and monitoring techniques that have not been covered by previous surveys in the field.

In the remainder of this paper, we describe the ageing mechanisms that affect embedded systems in  \autoref{sec:circuit-ageing}.
This includes the degradation of transistors, interconnects, and passive components.
\autoref{sec:survey} overviews the studied literature, introduces relevant taxonomies and the proposed classification system, and identifies trends in the techniques.
\autoref{sec:ageing-fpga}, \autoref{sec:ageing-mcu-soc}, and \autoref{sec:ageing-voltage} describe in detail ageing detection and monitoring strategies for FPGAs, microcontrollers and SoCs, and power supplies, respectively.
\autoref{sec:discussion} presents a discussion on possible research gaps in the field and an outlook with concrete future research directions.
The paper concludes in \autoref{sec:conclusion}.
 \section{Ageing Mechanisms in Embedded Systems}\label{sec:circuit-ageing}

A failure is a non-conformance to a defined performance criterion \cite{rmr-s}.
Dependable systems require reliability, which represents the probability of the system to stay failure-free for a given period of time (\ie to perform within specified limits).
Design requirements usually specify failure rates \(\lambda(t)\), which are the probability of failing at a certain point in time.

\begin{figure}[t]
    \centering
    \includegraphics[width=0.55\textwidth]{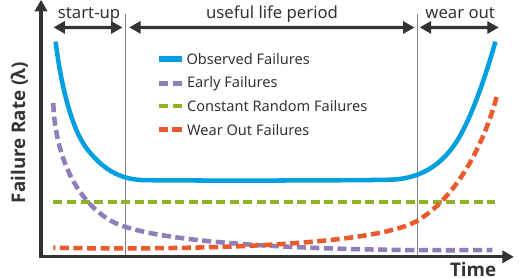}
    \caption{Bathtub distribution illustrating the typical evolution of the component failure rates over time \cite{rmr-s}.}
    \label{fig:bathtub}
\end{figure}

A component failure rate varies during its lifespan and is typically described using the bathtub distribution (see \autoref{fig:bathtub}).
The observed failures over time are the sum of three overlapping distributions, namely early failures, constant random failures, and wear-out failures, which lead to three distinct curve regions.
The first region is the start-up, also known as burn-in, in which the failure rate decreases over time.
Failures in this region are mostly due to manufacturing problems.
The middle region is the useful life period, in which failures occur randomly at a constant rate.
The last part of the distribution is the wear-out period.
The acceleration of degradation mechanisms (\eg component ageing) causes an increasing failure rate in this region.

Components of embedded systems suffer degradation during their lifetime.
Ageing processes are gradual, and their results are usually only noticeable in the long term.
Very Large-Scale Integration (VLSI) chips are subject to multiple wear-out or degradation processes that shift their parameters away from specification, eventually resulting in system failures \cite{damf-swc}.
Ageing in chips has two sources: transistor and interconnect degradation.
In addition, passive components such as capacitors and inductors age as well \cite{lpec-g}, causing problems in voltage regulation circuits \cite{itae-bhb}.

In this section, we describe the most relevant ageing mechanisms affecting the core components which compose devices commonly found in embedded systems.

\subsection{Ageing of Transistors}\label{transistor-ageing}

Silicon manufacturers encounter more challenges as they explore new miniaturized transistor technologies.
Efforts to further reduce device footprints prompted moving from planar to 3D topologies, such as FinFETs and gate-all-around FETs (GAAFETs).
This trend continues with vertical GAAFETs featuring vertical nanowires, which are predicted for usage beyond the \si{\qty{5}{\nano\meter}} scale \cite{nuhl-vhmj-19}.
These devices are susceptible to self heating, because the thermal paths to the ambient environment in their complex 3D structures are constrained.
An increase in operating temperature accelerates ageing processes and can reduce performance \cite{ispr-vs-19}.
Additionally, the reduction in size has not been proportionally followed by a lowered operational voltage, causing stronger electric fields applied to the devices.

Next, we describe the main ageing mechanisms that impact Metal-Oxide-Semiconductor Field-Effect Transistors (MOSFETs)--the base for all current CMOS technologies--, including newer technologies such as FinFETs.

\subsubsection{Bias Temperature Instability (BTI)}\label{bti}
BTI is a silicon ageing mechanism \cite{nbti-s, psbti-gkgr}, most prevalent on p-channel MOSFETs as Negative BTI (NBTI), which was initially observed on large technologies, such as \si{\qty{40}{\nano\meter}}.
Since the introduction of high-k/metal transistors, BTI effects are also considerable on n-channel MOSFETs, affected by Positive BTI (PBTI) \cite{rahd-swz}.
As devices scaled down, the necessity to boost doping levels accentuated the prominence of the random discrete dopant problem over NBTI \cite{drdm-smmn-02}.
Nevertheless, the ability to reduce doping levels due to the introduction of technologies like FinFETs, has resurfaced NBTI as the dominant time-dependent variability-inducing mechanism \cite{tshr-lkck-13, snbi-aga-19} which is actively under research \cite{enav-kmo-15, canv-pacm-19}.

NBTI induces an increase in the threshold voltage and a degradation of the carrier mobility, drain current, and transconductance on transistors.
Although the mechanism is not fully understood and many models are debated in research \cite{rnmm-mp-18}, the main cause for NBTI is attributed to the creation of traps in the oxide-substrate (\(SiO_2 / Si\)) interface and charges in the oxide.
The most prevalent model is the Reaction-Diffusion (R--D) \cite{nhfd-js, gdrm-os}.
It proposes that low-energy \(SiH\) bonds on the interface break due to the presence of applied electric field and high temperature, with a linear relation to the stress time (\ie reaction).
Afterwards, positive charges (\(H^+\) or trapped holes) diffuse into the gate oxide with a time dependence \(t^n\) (where usually \(n=1/4\)), leaving dangling bonds or interface traps behind \cite{nbti-s}.
The accumulation of positive charges in the oxide opposes the applied electric field, thus inducing a variation \(\Delta V_{th}\) in the threshold voltage of the transistor (\(V_{th}\)).
This reduces the maximum switching speed of the device.

BTI has a recovery phase: once the gate bias is removed, the diffused charges return to the interface and anneal the traps.
Although this process reduces \(\Delta V_{th}\), it is not completely clear whether the recovery is full:
some evidence indicates the possibility of a 100\% of recovery \cite{urbn-rmy}, while other studies propose the existence of both permanent and reversible degradation components \cite{drnb-eskw}.
As the mechanism is partly governed by a diffusion process, its duration highly relates to the operating temperature.
Owing to the recovery phase, degradation depends on the stress duty cycle; thus, continuous stress (DC) is much more severe than non-continuous (AC) \cite{nbti-sz}.

\subsubsection{Hot Carrier Injection (HCI)}\label{hci}

As a lateral electric field between drain and source is applied to a transistor, carriers (electrons or holes) gain kinetic energy, becoming \emph{``hot``} when their energy is significantly larger than the one of the lattice at thermal equilibrium \cite{rwmc-swvs}.
Due to their high energy, some carriers are able to surpass the potential barrier of the gate oxide, diffusing  into the dielectric or causing damage to the interface.
This has similar symptoms as BTI, namely an increase of threshold voltage \(\Delta V_{th}\), a decrease of carrier mobility, and a reduction of transconductance in the transistor saturation region \cite{heid-hthk, feth-ncdo, emdh-ts}, but unlike BTI, HCI presents no recovery phase when the stress is removed.
The impact of HCI is directly proportional to the frequency at which the device is switched, as hot carriers are generated during the transistor transitions.
HCI also depends on the device temperature, therefore its effects are exacerbated by the self heating process of thermally-complex technologies, such as FinFETs \cite{chcd-clcc-21}, for both p-MOS and n-MOS types \cite{ahcd-khsr-20, chcd-clcc-21}.
This situation is expected to worsen as 3D structures are built taller and narrower \cite{hcff-jlkk-16}, making it a clear challenge for technology scaling.

\subsubsection{Time-Dependent Dielectric Breakdown (TDDB)}\label{tddb}

TDDB is a gradual and irreversible degradation mechanism in the gate oxide of transistors.
Defects accumulated in the \(SiO_2\) layer allow an increase in gate leakage current, thus slowing down the transistor switching frequency.
This phenomenon has been studied on planar MOSFETs \cite{iact-rlnk-17} as well as on newer 3D FinFETs \cite{sstf-kjsk-18, refc-csjh-20}

\begin{figure}[t]
    \centering
    \includegraphics[width=0.50\textwidth]{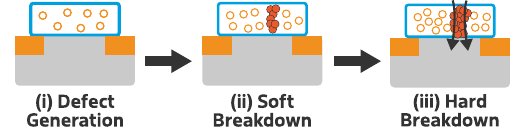}
    \caption{Cross-section of a transistor with the gate oxide traversing all three stages of dielectric breakdown.}
    \label{fig:tddb}
\end{figure}

Nigam \etal \cite{tsrc-nym} describe three stages of the degradation process as illustrated in \autoref{fig:tddb}:
\one Defect generation starts when a large electric field is applied to the gate dielectric.
As the defect density is still not critical, no gate  current leaks.
\two With continued stress, more defects are generated and start to overlap, producing a soft breakdown.
At this point, traps form a resistive path between the gate and the channel, increasing the leakage current and reducing the device switching speed.
\three Due to the thermal damage caused by the leakage current, more traps form a wider and less resistive path.
This increases the gate current even further, which leads to a thermal runaway that completely breaks the dielectric layer.
At this point, the transistor structure is destroyed, and the device is unusable.

\subsubsection{Radiation-Induced Trapped Charges}\label{tid}

When high-energy photons or charged particles interact with a material, they can cause ionization.
Total Ionizing Dose (TID) denotes the total amount of transferred energy from ionizing particles to the material.
As a consequence, semiconductors are vulnerable to photon-induced ionization damage via a process that generates two types of trapped charge: \one oxide-trapped charges and \two interface traps \cite{tidc-b}.
Trapped charges accumulate over time and modify the transistor characteristics.
The effect of radiation has been studied in many components commonly found in embedded systems: signal propagation changes within ICs \cite{stro-sbwm}, impact on voltage regulators \cite{rtco-nwgl}, System-On-Chips and microcontrollers \cite{tidc-kbol, igrm-srba, rtcm-fddn}.

The trap generation process initiates when an ionizing particle impacts the gate oxide material of the transistor, transferring part of its kinetic energy.
A number of electron-hole pairs are generated, proportional to the material activation energy and density \cite{admr-o}.
While some pairs are annihilated through recombination, which is a function of the applied electric field \cite{fdit-ssfw}, other pairs escape this process.
A fraction of the remaining holes may fall into deep traps in the oxide or near the \(Si/SiO_2\) interface, forming the trapped positive charges.
Other holes react with hydrogen-containing defects in the interface, generating interface traps.

\begin{figure}[t]
    \centering
    \includegraphics[width=0.50\textwidth]{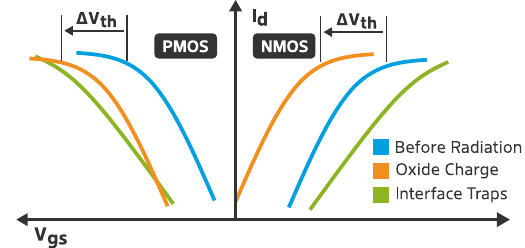}
    \caption{Illustration of the effect that fixed oxide trapped charges have on the drain current (\(I_d\)) \vs gate-source voltage (\(V_{gs}\)) characteristic for N-MOS and P-MOS devices \cite{tidc-b}.}
    \label{fig:tid-effects}
\end{figure}

Oxide traps modify the DC characteristics of CMOS circuits, similarly to BTI, as depicted in \autoref{fig:tid-effects}.
The most prominent effect is a negative shift in the drain current \(I_d\) for a given gate-source voltage \(V_{gs}\) on P-MOS and N-MOS.
In the first one, the absolute value of \(V_{th}\) increases while the drain current is reduced.
The latter one suffers a reduction of \(V_{th}\), with an increase of the drive current, potentially causing a latch-up (\ie the transistor always conducts).
Interface traps affect the recombination rates of carriers and their mobility through the channel.
As the trapping and de-trapping of charges at the interface depend on the applied bias voltage, when these traps build up, they generate an increase in the sub-threshold swing of CMOS devices.
The consequence is that the \(I_d\) vs.~\(V_{gs}\) response is stretched out, as shown in \autoref{fig:tid-effects}.

Recent research on modern transistor technologies has shown an increased TID tolerance on \si{\qty{28}{\nano\meter}} MOSFETs and \si{\qty{16}{\nano\meter}} FinFETs, when compared to previous \si{\qty{65}{\nano\meter}} nodes \cite{tmfh-bmmb-22}.
In the case of FinFETs, the TID tolerance was found to strongly depend on the channel length and the number of fins on the transistor \cite{iftid-mbmb-22}.
Furthermore, Gorchichko \etal \cite{tidr-gzwb-21} measured an extremely high TID tolerance for GAAFETs, which is partly due to thin gate dielectrics that prevent building up charges and make them good candidates for radiation-enhanced applications.

\subsection{Ageing of Interconnects}\label{interconnects-ageing}

\subsubsection{Electromigration}\label{em}

To keep up with Moore's Law \cite{ccic-m}, manufacturers constantly increase levels of integration.
One way to integrate more devices is to reduce the width of the wires interconnecting them.
This optimization comes at a cost: the smaller the conducting area, the higher the current density.
Increasing current density entails a higher atomic diffusion phenomenon called electromigration.
In the presence of high current densities, the strong momentum transfers from electrons to the conductor atoms to cause an atomic diffusion in the direction of the electron flow \cite{emic-tl}.
The force of the electronic wind causes atoms to deplete ``up wind`` and accumulate ``down wind``, until they create an electrical short- or open-circuit, rendering ICs unusable.
Black~\cite{emsr-b} proposed to model the Median Time to Failure (MTF) in hours by \autoref{eq:mtf-electromigration}, where \(A\) is a constant, \(J\) is the current density, \(\phi\) is the activation energy for diffusion, \(k_\textup{B}\) is the Boltzmann constant, and \(T\) is the temperature in Kelvin.

\begin{equation}
	MTF = A J^{-2} \exp{\left({\frac{\phi}{k_\textup{B} T}}\right)}
    \label{eq:mtf-electromigration}
\end{equation}

\subsection{Ageing of Passive Components}

\subsubsection{Capacitors}\label{subsec:ageing-caps}

Even though it is well known that electrolytic capacitors wear out with time \cite{lpec-g}, they are widely used in many embedded applications, from filtering signals to suppressing voltage ripples.
As digital circuits lower their operating voltage, they become more susceptible to noise and require functioning capacitors over the whole deployment time.
Most of the models representing real capacitors in electrical circuits describe an Equivalent Series Resistance (ESR), which causes energy dissipation as heat.
Although ideally the ESR value is negligible, changes have been observed over time when the component is exposed to high temperatures \cite{pelc-g}.
The vaporization of the capacitor electrolyte through the encapsulation seal increases the ESR.
As vaporization depends on the internal temperature and the ESR dissipates heat, it acts in positive feedback degrading the capacitor parameters.

\subsubsection{Inductors}\label{subsec:ageing-inductors}

Inductors are commonly employed to build analog filters and power supplies.
Magnetic cores are usually inserted inside the coils to increase inductance.
Depending on their composition and geometry, ferromagnetic cores generate energy losses due to eddy currents in the material induced by the fluctuating magnetic field.
Indeed, according to Faraday's law of induction, upon the presence of a changing magnetic flux \(\boldsymbol{\Phi}\) in an enclosed path, an opposite electromotive force \(\boldsymbol{\mathcal{E}}\) is induced.
This force will be proportional to the rate of change of the flux, as \autoref{eq:faraday} shows.

\begin{equation}
    \boldsymbol{\mathcal{E}} = - \frac{d}{dt} \boldsymbol{\Phi}
    \label{eq:faraday}
\end{equation}

The electric flux crossing a surface \(\Sigma\) (bounded by a path \(C\)) can be expressed as \autoref{eq:flux}, where \(\boldsymbol{{B}}\) is the magnetic field.

\begin{equation}
    \boldsymbol{\Phi} = \iint_{\Sigma} {\boldsymbol{B} \cdot dA}
    \label{eq:flux}
\end{equation}

The induced \(\boldsymbol{\mathcal{E}}\) relates to the generated electric field \(\boldsymbol{E}\) over the closed path \(C\), as shown in \autoref{eq:electric-field}.

\begin{equation}
    \boldsymbol{\mathcal{E}} = \oint_C \boldsymbol{E} \cdot dl
    \label{eq:electric-field}
\end{equation}

Finally, \autoref{eq:electric-magnetic-fields} shows that the varying magnetic field in the enclosed inductor core will induce an electric field in the ferromagnetic material.
Whenever the material of the core is conducting, a current density will appear on it.

\begin{equation}
    \oint_C \boldsymbol{E} \cdot dl = - \frac{d}{dt} \iint_{\Sigma} {\boldsymbol{B} \cdot dA}
    \label{eq:electric-magnetic-fields}
\end{equation}

The power dissipation of the eddy currents depends on the induced electric field on the core and its resistivity.
Inductor cores can undergo a degradation process when constantly exposed to elevated temperatures for long periods of time, which changes the material resistivity.
This increases core losses \cite{itae-bhb} and contributes to further elevating the temperature.
 \section{Surveying Ageing Detection Techniques}\label{sec:survey}

A defect is an unintended deviation of a component material of physical nature and persistent effects \cite{stsa-kw}.
Defects originate either from manufacturing imperfections or from ageing during service.
Defective behaviour can be abstracted to faults when modelled according to the circuit structure.
For instance, a transistor threshold voltage degradation (\ie a physical variation) can be modelled as a delay fault on a combinational circuit.
When the system state or its environment triggers a fault, an error occurs.
This may, in turn, cause a malfunction or complete failure.

The variety of potential defects of electronic components (\cf \autoref{sec:circuit-ageing}) evidences the need for health assessment of embedded hardware.
Prognostics and remaining-useful-life estimations \cite{ieee-1856-2017} can leverage this information.
The evolution of physical ageing processes can indirectly be assessed during operation by observing deviations of device parameters, either by directly comparing against pre-established profiles or with sophisticated anomaly detection techniques \cite{eada-mgbe}.
This opens the door to fully ageing-aware embedded systems that  monitor, analyse, and act upon low-level degradation information, either by means of mitigation \cite{camt-kadk-17} or even by self-healing \cite{fdsh-gs-20} techniques.

\subsection{Surveying Methodology}\label{subsec:methodology}

The research for the presented literature comprised consulting of various notable scientific databases (IEEE, ACM, Scopus, and Google Scholar) while limiting the results with keywords including \emph{ag(e)ing, degradation, failure, monitoring}, and the components of interest: FPGA, microcontroller, system on chip, and power supply.
We excluded, however, board-level mechanical wear-out effects, such as broken interconnects or the degradation of solder joints.
Upon the initial searches, we followed the related work referenced by the results.
References were selected with the objective of maximizing the variety of detection and monitoring approaches.
We did not include work on the design of special ageing hardware probes for ASICs, instead we focused on techniques applied to COTS devices without built-in on-chip ageing sensors.
Moreover, the presented techniques have been implemented in real hardware; therefore, their results are empirical and not exclusively based on simulations.
This literature compendium includes methods originally designed to assess the system health status while on service, as well as methods that have been employed to study the impact of ageing processes on devices under controlled environments.

We consider that the selected components represent well the minimal building blocks present in most modern embedded systems: a main processing unit, memory, and a power source \cite{esd-m-21}.
While a huge variety of peripherals, such as sensors and actuators, are frequently required on certain embedded system domains (\eg cyber-physical systems and IoT \cite{wtgo-gvgs-18}), they are use-case specific in nature and can vary fundamentally in their functioning principles.
We aim to abstract from particular deployment cases for embedded devices.
Instead, we aspire for this survey to serve as a comprehensive guide across embedded domains and thus focus on the body of work concerning FPGAs, MCUs, SoCs, and power supplies.

\subsection{Classification of Tests}\label{subsec:classification-of-tests}

\begin{figure}[t]
    \centering
    \includegraphics[width=0.8\textwidth]{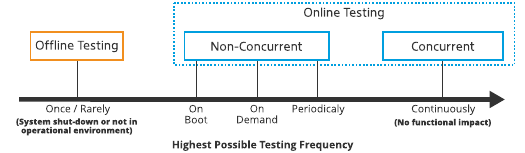}
    \caption{Test types organized according to their highest testing frequency. We have expanded the classification system proposed by Kochte \etal \cite{stsa-kw}, which focuses on self-testing systems, in order to apply it to ageing detection techniques.}
    \label{fig:test_classification}
\end{figure}

Given the wide variety of tests that can be performed on the systems under consideration, possible classifications vary depending on the intended use case.
In this work, we adapted the classification system proposed by Kochte \etal \cite{stsa-kw} and extended it as presented in \autoref{fig:test_classification}.

The primary variable is the highest testing frequency that each technique can potentially achieve.
The classification distinguishes between infrequent testing that requires a system shut-down or special testing equipment, \ie \emph{offline testing}, and testing that can be performed autonomously onboard, \ie \emph{online testing} or self-testing.
Moreover, online testing can either be \emph{concurrent} or \emph{non-concurrent} depending on whether it runs without influencing the system operation.

\subsubsection{Offline Testing}\label{subsec:ext-tests}

Offline testing usually involves automatic testing equipment and extra hardware or tools.
These tests are typically required whenever the observed magnitudes need special detection techniques or sensors not present in the system (\eg radiated electromagnetic emissions).
Offline tests tend to be much more expensive and cumbersome to carry out than online tests.
For this reason, they are mainly reserved for in-lab evaluations under controlled environments and are rarely deployed in the field.
Potential applications of offline tests are acceptance testing for quality assurance (\eg to detect counterfeits) and post-mortem analysis.

\subsubsection{Online Testing}\label{subsec:self-test}

Online tests or self-tests leverage the existing capabilities of the devices under test to evaluate their own functionality, thus avoiding extra equipment.
Depending on their measurement methodology, online tests are \one concurrent or \two non-concurrent.
The former run in parallel with the applications and are useful to monitor systems that cannot afford downtime.
The latter are performed outside the device service schedule (\eg during boot or maintenance periods).

\paragraphc{Concurrent Online Testing}\label{subsec:conc-test}~is a non-intrusive testing approach that requires no suspension of the system service.
Such techniques are ideal for ageing detection in safety-critical deployments that operate without interruption for extended periods of time, \eg nuclear plants or large experimental facilities.

\paragraphc{Non-Concurrent Online Testing}\label{subsec:nonconc-test}~consists of tests that cannot be performed on a running system due to their intrusiveness, \eg the test requires a reconfiguration in order to be performed.
Ageing detection using non-concurrent online tests is possible for systems that can be periodically rebooted or reconfigured.

\subsection{Overview of the Results}\label{subsec:overview}
Considering the criteria described in \autoref{subsec:methodology} and the taxonomy introduced in \autoref{subsec:classification-of-tests}, relevant literature has been selected, reviewed, and classified.
\autoref{table:summary} summarizes this related work, which is discussed in detail in the following sections.

Approaches are organized by component: FPGA, Microcontroller Unit (MCU) and SoC, and Low Drop-Out (LDO) and Switching-Mode Power Supply (SMPS).
Contributions are further grouped according to their sensing principle.
Within each group, references are chronologically ordered, so that the latest appear at the end.
For each publication we indicate whether the technique was developed as an online concurrent, online non-concurrent, or offline test.
Moreover, we expose whether techniques were evaluated under aged conditions, whether machine learning (ML) techniques were applied, and whether an analytical or empirical model was developed to study the degradation process.

\begin{center}
   \small
\begin{longtable}[t]{|c|c?l|c|c|c|c|c|c|}
   \caption{
      Summary of ageing detection and monitoring techniques grouped by component and sensing principle.
      Within each sensing principle, references are chronologically organized so that newer ones are at the bottom.
      For each reference it is indicated the type of test (concurrent, non-concurrent, or offline), whether machine learning models have been applied, and whether a model of the degradation processes has been developed.
      Additionally, it is shown whether the technique has been evaluated under electrical stress (\faBolt), thermal stress (\faThermometerThreeQuarters), radiation (\faRadiation*), natural ageing (\faHourglassHalf), or a combination of them.
   } \label{table:summary} \\

   \hline
   {\multirow{2}{*}{\textbf{Component}}} & \multirow{2}{*}{\textbf{\makecell[c]{Sensing\\Principle}}} & \multicolumn{1}{c|}{\multirow{2}{*}{\textbf{Reference}}} & \multicolumn{2}{c|}{\textbf{Online}} & \multirow{2}{*}{\textbf{\makecell[c]{Offline}}} & \multirow{2}{*}{\textbf{\makecell[c]{Ageing}}} & \multirow{2}{*}{\textbf{\makecell[c]{ML}}} & \multirow{2}{*}{\textbf{\makecell[c]{Model}}} \\
   \cline{4-5}
    & & & Conc. & Non-conc. & & & &\\
   \hline \hline
   \endfirsthead

   \multicolumn{9}{c}{{\bfseries \tablename \thetable{} -- continued from previous page}} \\
   \hline
   {\multirow{2}{*}{\textbf{Component}}} & \multirow{2}{*}{\textbf{\makecell[c]{Sensing\\Principle}}} & \multicolumn{1}{c|}{\multirow{2}{*}{\textbf{Reference}}} & \multicolumn{2}{c|}{\textbf{Online}} & \multirow{2}{*}{\textbf{\makecell[c]{Offline}}} & \multirow{2}{*}{\textbf{\makecell[c]{Ageing}}} & \multirow{2}{*}{\textbf{\makecell[c]{ML}}} & \multirow{2}{*}{\textbf{\makecell[c]{Model}}} \\
   \cline{4-5}
    & & & Conc. & Non-conc. & & & &\\
   \hline \hline
   \endhead

   \hline \multicolumn{9}{|r|}{{Continued on next page}} \\ \hline
   \endfoot

   \hline
   \endlastfoot

   \multirow{30}{*}{\begin{minipage}{5em}\centering\textbf{FPGA}\\(\autoref{sec:ageing-fpga})\end{minipage}} & \multirow{10}{*}{\begin{minipage}{6em}\centering Shadow Register \end{minipage}} & {\footnotesize \TBstrut Li \cite{nsrd-ll}} & {\footnotesize \faCheck} & & & & &\\
   \cline{3-9}
    & & {\footnotesize \TBstrut Wong \cite{sccd-wsc}} & & {\footnotesize \faCheck} & & & & \\
   \cline{3-9}
    & & {\footnotesize \TBstrut Wong \cite{smcd-wsc}} & & {\footnotesize \faCheck} & & & & \\
   \cline{3-9}
    & & {\footnotesize \TBstrut Valdes \cite{psoc-vfmr}} & {\footnotesize \faCheck} & & & & & \\
   \cline{3-9}
    & & {\footnotesize \TBstrut Amouri \cite{alsa-at}} & {\footnotesize \faCheck} & & & & & \\
   \cline{3-9}
    & & {\footnotesize \TBstrut Pfeifer \cite{drxd-pp}} & {\footnotesize \faCheck} & & & & & \\
   \cline{3-9}
    & & {\footnotesize \TBstrut Leong \cite{amls-lsts}} & {\footnotesize \faCheck} & & & & & \\
   \cline{3-9}
    & & {\footnotesize \TBstrut Valdes \cite{dvcs-vfmr}} & {\footnotesize \faCheck} & & & & & \\
   \cline{3-9}
    & & {\footnotesize \TBstrut Ghaderi \cite{hssd-genb}} & {\footnotesize \faCheck} & & & & & \\
   \cline{3-9}
    & & {\footnotesize \TBstrut Jiang \cite{afdm-jlth}} & & {\footnotesize \faCheck} & & & & \\

   \cline{2-9}
    & \multirow{17}{*}{\begin{minipage}{4em} \centering Ring Oscillator \end{minipage}} & {\footnotesize \TBstrut Ruffoni \cite{dmpd-rb}} & & & {\footnotesize \faCheck} & & & \\
   \cline{3-9}
    & & {\footnotesize \TBstrut Sedcole \cite{wdvf-sc}} & & {\footnotesize \faCheck} & & & & {\footnotesize \faCheck} \\
   \cline{3-9}
    & & {\footnotesize \TBstrut Zick \cite{oshf-zh}} & {\footnotesize \faCheck} & & & & & {\footnotesize \faCheck} \\
   \cline{3-9}
    & & {\footnotesize \TBstrut Bruguier \cite{pcfe-bbpm}} & & & {\footnotesize \faCheck} & & & \\
   \cline{3-9}
    & & {\footnotesize \TBstrut Pfeifer \cite{mimf-pp}} & & {\footnotesize \faCheck} & & & & \\
   \cline{3-9}
    & & {\footnotesize \TBstrut Pfeifer \cite{mppm-pp}} & & {\footnotesize \faCheck} & & {\footnotesize \faThermometerThreeQuarters} & & \\
   \cline{3-9}
   & & {\footnotesize \TBstrut Amouri \cite{aefa-abkb}} & & & {\footnotesize \faCheck} & {\footnotesize {\faBolt}, {\faThermometerThreeQuarters}} & & \\
   \cline{3-9}
   & & {\footnotesize \TBstrut Pfeifer \cite{rloc-pkp}} & & {\footnotesize \faCheck} & & {\footnotesize \faHourglassHalf} & & \\
  \cline{3-9}
    & & {\footnotesize \TBstrut Naouss \cite{ditb-nm}} & & {\footnotesize \faCheck} & & {\footnotesize \faBolt} & & \\
   \cline{3-9}
    & & {\footnotesize \TBstrut Naouss \cite{mddf-nm}} & & {\footnotesize \faCheck} & & {\footnotesize \faBolt} & & {\footnotesize \faCheck} \\
   \cline{3-9}
    & & {\footnotesize \TBstrut Naouss \cite{fldd-nm}} & & {\footnotesize \faCheck} & & {\footnotesize \faBolt} & & {\footnotesize \faCheck} \\
   \cline{3-9}
    & & {\footnotesize \TBstrut Maragos \cite{apvf-mtls}} & & {\footnotesize \faCheck} & & & & \\
   \cline{3-9}
    & & {\footnotesize \TBstrut Bender \cite{rpff-bbb-20}} & & {\footnotesize \faCheck} & & {\footnotesize {\faBolt}, {\faThermometerThreeQuarters}} & & {\footnotesize \faCheck} \\
   \cline{3-9}
    & & {\footnotesize \TBstrut Ahmed \cite{arfd-asi}} & & & {\footnotesize \faCheck} & {\footnotesize \faThermometerThreeQuarters} & {\footnotesize \faCheck} & {\footnotesize \faCheck} \\
   \cline{3-9}
    & & {\footnotesize \TBstrut Li \cite{iama-lhww-22}} & & {\footnotesize \faCheck} & & {\footnotesize {\faBolt}, {\faThermometerThreeQuarters}} & {\footnotesize \faCheck} & {\footnotesize \faCheck} \\
   \cline{3-9}
    & & {\footnotesize \TBstrut Sobas \cite{dmma-sm-24}} & & {\footnotesize \faCheck} & & {\footnotesize {\faBolt}, {\faThermometerThreeQuarters}, {\faHourglassHalf}} & & {\footnotesize \faCheck} \\
   \cline{3-9}
    & & {\footnotesize \TBstrut Lanzieri \cite{lbkfs-sdpdf-24}} & & {\footnotesize \faCheck} & & {\footnotesize {\faHourglassHalf, \faRadiation*}} & {\footnotesize \faCheck} & \\
   \cline{2-9}

   \cline{2-9}
    & \multirow{3}{*}{\begin{minipage}{5em} \centering Transition Probability\end{minipage}} & {\footnotesize \TBstrut Wong \cite{tpdm-wdc}} & & {\footnotesize \faCheck} & & & & \\
   \cline{3-9}
    & & {\footnotesize \TBstrut Stott \cite{damf-swc}} & & {\footnotesize \faCheck} & & & & \\
   \cline{3-9}
    & & {\footnotesize \TBstrut Stott \cite{dfmm-swsc}} & & {\footnotesize \faCheck} & & {\footnotesize {\faBolt}, {\faThermometerThreeQuarters}} & & {\footnotesize \faCheck} \\

   \hline \hline
   \multirow{4}{*}{\begin{minipage}{5em}\centering\textbf{MCU \& SoC}\\(\autoref{sec:ageing-mcu-soc})\end{minipage}} & \multirow{4}{*}{\begin{minipage}{4em}\centering SRAM Pattern \end{minipage}} & {\footnotesize \TBstrut Guo \cite{zprs-grtf}} & & {\footnotesize \faCheck} & & {\footnotesize {\faBolt}, {\faThermometerThreeQuarters}} & & \\
   \cline{3-9}
    & & {\footnotesize \TBstrut Guo \cite{scar-gxrt}} & & {\footnotesize \faCheck} & & {\footnotesize {\faBolt}, {\faThermometerThreeQuarters}} & & \\
   \cline{3-9}
    & & {\footnotesize \TBstrut Guin \cite{drsa-gwhs}} & & {\footnotesize \faCheck} & & {\footnotesize \faThermometerThreeQuarters} & & \\
   \cline{3-9}
    & & {\footnotesize \TBstrut Lanzieri \cite{lkfss-aaesl-23}} & & {\footnotesize \faCheck} & & {\footnotesize \faHourglassHalf} & {\footnotesize \faCheck} & \\
   \cline{2-9}

   \multirow{6}{*}{\begin{minipage}{5em}\centering\textbf{MCU \& SoC}\\(\autoref{sec:ageing-mcu-soc})\end{minipage}} &  Time Window & {\footnotesize \TBstrut Diggins \cite{tidi-dmhk}} & & & {\footnotesize \faCheck} & {\footnotesize \faRadiation*} & & \\
   \cline{2-9}

    & \multirow{2}{*}{\begin{minipage}{4em} \centering LDO PSRR \end{minipage}} & {\footnotesize \TBstrut Chowdhury \cite{rsdl-cgf}} & & & {\footnotesize \faCheck} & {\footnotesize \faThermometerThreeQuarters} & {\footnotesize \faCheck} & \\
   \cline{3-9}
    & & {\footnotesize \TBstrut Acharya \cite{lbor-avf}} & & & {\footnotesize \faCheck} & & {\footnotesize \faCheck} & \\
   \cline{2-9}
    & \multirow{3}{*}{\makecell[c]{EM Charact.}} & {\footnotesize \TBstrut Dawson \cite{ehtm-dfdm}} & & & {\footnotesize \faCheck} & {\footnotesize \faThermometerThreeQuarters} & & \\
   \cline{3-9}
    & & {\footnotesize \TBstrut Li \cite{cmsa-lwhz}} & & & {\footnotesize \faCheck} & {\footnotesize \faThermometerThreeQuarters} & & \\
   \cline{3-9}
    & & {\footnotesize \TBstrut Wu \cite{msaa-wllz}} & & & {\footnotesize \faCheck} & {\footnotesize {\faBolt}, {\faThermometerThreeQuarters}} & & \\
   \cline{2-9}

   \hline \hline
   \multirow{9}{*}{\begin{minipage}{5em}\centering\textbf{Power \\Supply}\\(\autoref{sec:ageing-voltage})\end{minipage}} & \multirow{4}{*}{\begin{minipage}{5em}\centering Equivalent Series Resistance \end{minipage}} & {\footnotesize \TBstrut Lahyani \cite{fpec-lvgv}} & {\footnotesize \faCheck} & & & {\footnotesize \faThermometerThreeQuarters} & & {\footnotesize \faCheck} \\
   \cline{3-9}
   & & {\footnotesize \TBstrut Chen \cite{ecfp-ccw}} & {\footnotesize \faCheck} & & & {\footnotesize \faThermometerThreeQuarters} & & \\
   \cline{3-9}
   & & {\footnotesize \TBstrut Boyer \cite{itae-bhb}} & & & {\footnotesize \faCheck} & {\footnotesize \faThermometerThreeQuarters} & & {\footnotesize \faCheck} \\
   \cline{3-9}
   & & {\footnotesize \TBstrut Givi \cite{cmdc-gfg}} & {\footnotesize \faCheck} & & & & & \\
   \cline{2-9}

   & \multirow{3}{*}{\makecell[c]{EM Charact.}} & {\footnotesize \TBstrut Wu \cite{esvr-wlsb}} & & & {\footnotesize \faCheck} & {\footnotesize \faBolt} & {\footnotesize \faCheck} & \\
   \cline{3-9}
   & & {\footnotesize \TBstrut Boyer \cite{itae-bhb}} & & & {\footnotesize \faCheck} & {\footnotesize \faThermometerThreeQuarters} & & {\footnotesize \faCheck} \\
   \cline{3-9}
   & & {\footnotesize \TBstrut Boyer \cite{stab-bggd}} & & & {\footnotesize \faCheck} & {\footnotesize \faThermometerThreeQuarters} & & {\footnotesize \faCheck} \\
   \cline{2-9}
   & \multirow{2}{*}{PSRR} & {\footnotesize \TBstrut Chowdhury \cite{aldo-cspm}} & & & {\footnotesize \faCheck} & {\footnotesize {\faBolt}, {\faThermometerThreeQuarters}} & & \\
   \cline{3-9}
   & & {\footnotesize \TBstrut Chowdhury \cite{rcdl-cgbm}} & & & {\footnotesize \faCheck} & {\footnotesize \faThermometerThreeQuarters} & {\footnotesize \faCheck} & \\
   \cline{2-9}

   \hline
\end{longtable}
\end{center}

At first glance, \autoref{table:summary} reveals that most of the literature focusing on FPGAs proposes online test approaches, while related work on MCUs, SoCs, or power supplies dominantly relies on offline tests using external equipment.
While various publications in the field of FPGAs and power supplies develop analytical and empirical models of the ageing mechanisms at play, this is not the case for MCUs and SoCs.
In FPGA studies, ring oscillators are the preferred technique for characterizing and modelling.
Finally, we can also observe a trend towards the application of ML in recent years, mainly for ageing detection techniques.

\begin{figure}[t]
    \centering
    \includegraphics[width=0.95\textwidth]{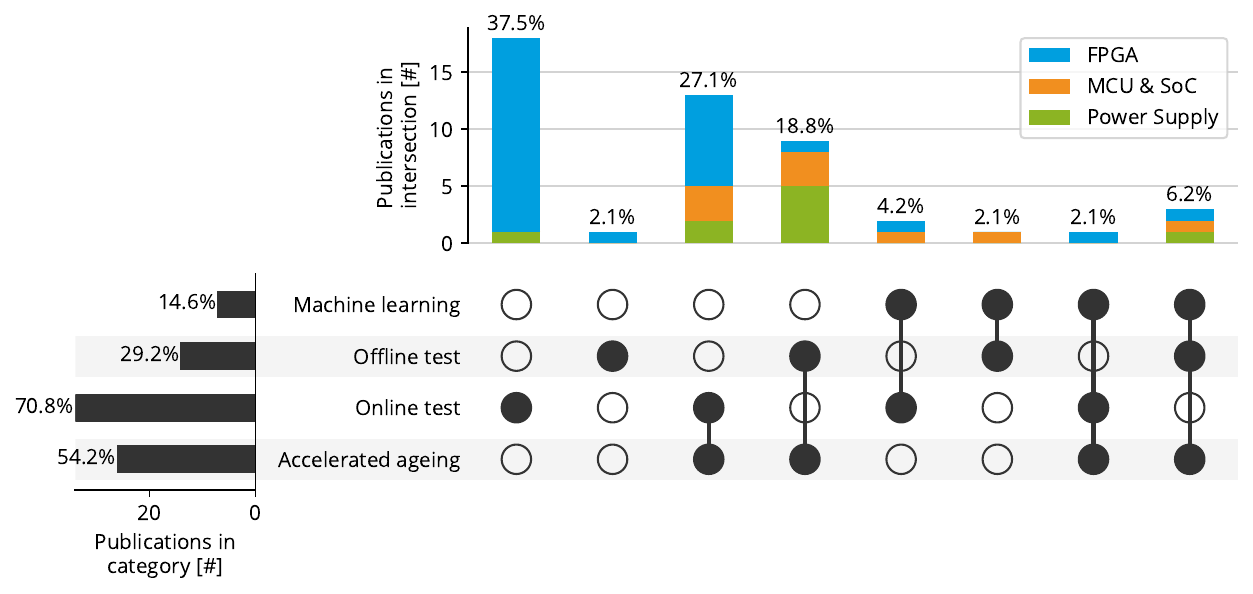}
    \caption{The UpSet plot visualizes exclusive intersections between publications in different categories.
        The left bars show the total number of references that belong to each category and the percentage of the total.
        The upper bars illustrate how many of the publications belong to each exclusive intersection of categories, which are indicated in the matrix below.
        Each intersection (\ie each vertical bar) includes papers in the highlighted categories (\desyPetrolCircle{\phantom{c}}) and excludes the ones in the non-highlighted categories (\desyPetrolCircleEmpty{\phantom{c}}).
    }
    \label{fig:upset}
\end{figure}

By observing the intersections of categories, we can further characterize the  state-of-the-art.
The UpSet plot \cite{upset-lgsv-14} in \autoref{fig:upset} shows the various shares of publications that fit into one or several classification categories.
Even though \si{\qty{70}{\percent}} of the techniques are based on online tests, this number reduces to \si{\qty{27}{\percent}} if we restrict to those evaluated under accelerated ageing conditions, of which more than half apply to FPGAs.
This differs for techniques describing offline tests, of which more than \si{\qty{85}{\percent}} are tested using accelerated ageing.
Notably, more than \si{\qty{57}{\percent}} of the techniques that involve ML models require offline tests to work.

 \section{Ageing Detection on FPGAs}\label{sec:ageing-fpga}

FPGAs are programmable hardware devices that provide great flexibility, in-field updates, and rapid prototyping of digital designs.
The majority of FPGA applications require high-speed processing, including digital filters and control loops with strict low latency requirements.
SRAM-based FPGAs are the most popular in the market and use volatile memory to store their configuration.
These devices present a regular structure of configurable logic blocks and programmable switch matrices that allow for a mapping of logic functions.

FPGAs face reliability challenges as their transistor-based circuits are affected by ageing mechanisms.
BTI and HCI increase signal propagation delays \cite{aefa-abkb}.
With the gradual decay of delays in the signals, critical paths of digital applications may change and even reach values out of specification.

Given their general purpose nature, and unlike ASICs, manufacturers cannot know beforehand what functionality will be synthesized on the FPGA chip, hindering the implementation of precautionary measures to cope with circuit degradation (\eg guard-bands).
This situation moves the issue of monitoring ageing processes to the application developers, who are required to know the tolerances of signal propagation times for their specific use case.

\subsection{Ring Oscillator} \label{subsec:fpga-ro}

\begin{figure}[t]
    \centering
    \includegraphics[width=0.55\textwidth]{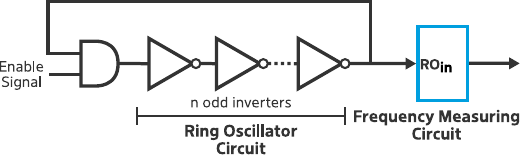}
    \caption{Generic sensor based on a ring oscillator with an enable signal, see also \cite{bdt-iaroc-17}.}
    \label{fig:ro-sensor}
\end{figure}

The flexibility of FPGAs enables the implementation of monitoring techniques directly on the programmable hardware.
 Ring Oscillators (ROs), which consist of an odd number of inverters connected in a ring, are one of the most common approaches to measuring digital signal propagation time.
RO-based sensors are usually composed of a RO and a frequency-measuring circuit (see \autoref{fig:ro-sensor}).
The frequency at the output of a RO depends -- at a given temperature and voltage -- on the inverter propagation times \(t_p\) and the number of gates \(n\), as shown in \autoref{eq:ro-freq}.

\begin{equation}
    f = \frac{1}{2 \cdot n \cdot t_p}
    \label{eq:ro-freq}
\end{equation}

Zick \etal \cite{oshf-zh} proposed an online concurrent sensing method to measure variations in physical parameters.
By using an enhanced RO, an efficient counter, and control logic, the authors developed a compact sensor requiring only 8 Look-Up Tables (LUTs).
A residue number system ring counter was implemented, as it requires fewer resources than a binary counter.
The temperature sensitivity of the RO was increased to detect hotspots on the chip.
RO sensors were placed regularly on a hexagonal tessellation all over the FPGA, together with a softcore, a timer, and a UART.
The authors measured propagation times and indirectly estimated a transistor current leakage profile, localized dynamic power usage and temperature.

Sedcole \etal \cite{wdvf-sc} placed ROs in a matrix configuration to measure within-die delay variability.
The structure enabled the encoding of regions under test into columns and rows.
In such a dense configuration, ROs should only be active for short periods of time to avoid heating neighbouring sensors.
A single counter, timer, and unified control logic performed the measurements.
Pfeifer \etal characterized chip delays during design time \cite{mimf-pp} and analysed the limitations of 28 nm FPGAs \cite{mppm-pp} with ROs.
To measure the oscillations, they implemented a method based on Block RAM (BRAM), which was later formalized in an approach called ``Reliability-on-Chip`` \cite{rloc-pkp}.
The approach requires a true-dual-port BRAM on the chip to undersample the oscillator outputs and a softcore to read the signal streams for calculating the delays.
Li \etal \cite{iama-lhww-22} not only studied the process variability of \si{\qty{28}{\nano\meter}} FPGAs under NBTI by deploying ROs, but also utilized the data extracted from accelerated ageing to train various ML models.
By building datasets with information regarding the artificial ageing stress conditions and the RO sensor configurations, they found that the XGBoost model performed  best across all conditions.

Lanzieri \etal \cite{lbkfs-sdpdf-24} developed a RO measurement module to perform a large-scale study of the propagation delay on 298 Xilinx Virtex-6 devices, which have been naturally aged and in operation as part of a linear particle accelerator.
By comparing delay measurements of used and unused devices, the authors found evidence of effects caused by ageing mechanisms.
Moreover, an analysis of the radiation exposure of the devices inside the accelerator showed that slower propagation delays are correlated to higher radiation doses.

The application of ROs as variability sensors extends to more modern node technologies, as well.
Maragos \etal \cite{apvf-mtls} explored increased intra- and inter-die variability on \si{\qty{16}{\nano\meter}} FinFET FPGAs with various ROs controlled by an embedded Cortex-A53 CPU.
After an ageing process of \si{\qty{8000}{\hour}}, Sobas \etal \cite{dmma-sm-24}  measured the degradation of various \si{\qty{16}{\nano\meter}} FinFET FPGAs by implementing an RO-based test bench.
The authors evaluated the effects of static and dynamic stress, and derived an empirical degradation model for both modes that yields estimations with less than \si{\qty{10}{\percent}} relative error.
When comparing their results to similar evaluations on \si{\qty{28}{\nano\meter}} MOSFETs, they found that the smaller nodes show better reliability on static stress, and that BTI was the predominant ageing mechanism (as opposed to HCI).
A similar conclusion was reached by Bender \etal \cite{rpff-bbb-20} from assessing \si{\qty{16}{\nano\meter}} FinFET Xilinx devices with a multi-temperature operational life testing method.
By evaluating the impact of different temperatures, voltages, frequencies, and RO sizes, the authors managed to isolate the effects of various degradation mechanisms.
Additionally, their results suggest that there is a contribution from the self-heating effect to BTI, due to the lower heat dissipation of the transistor fins.

\begin{figure}[t]
    \centering
    \includegraphics[width=0.55\textwidth]{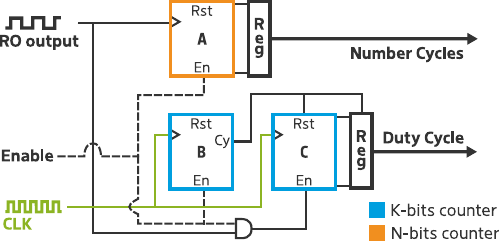}
    \caption{Measuring circuit of the test bench proposed in \cite{ditb-nm}, which counts the number of cycles of the RO signal and its duty cycle.}
    \label{fig:duty-cycle-measuring}
\end{figure}

Naouss \etal proposed \cite{ditb-nm} a test bench to self-characterize the delay of LUTs.
Their design allows stress signals to be injected into the Circuit Under Test (CUT) to produce accelerated ageing.
This feature was later used to independently study BTI \cite{mddf-nm} and HCI \cite{fldd-nm} impact by using different stress signals.
They implemented a frequency measuring circuit using three asynchronous counters (one of N bits and two of K bits) and a clock reference.
This circuit (see \autoref{fig:duty-cycle-measuring}) allowed for counting the number of cycles in a given period of time as well as the duty cycle of the signal.
An enable signal activated the counters and controlled the counting window.
Counter A registered the number of cycles of the RO output signal, which was used to calculate the oscillation frequency.
Counter B counted the number of clock cycles during the active time of the RO signal, from which the duty cycle was calculated.
Measuring the signal duty cycle allowed for studying the impact of ageing on the rising and falling times.

ROs have been employed in offline tests as well \cite{arfd-asi, dmpd-rb, aefa-abkb, pcfe-bbpm}.
Ahmed \etal \cite{arfd-asi} used performance degradation to detect recycled FPGAs by exhaustively fingerprinting LUT delays.
They synthesized ROs with routes configurable via an SRAM.
External equipment was used to control the test logic and read out the frequency of each oscillator.
Ruffoni \etal \cite{dmpd-rb} focused on measuring delays of the internal FPGA wires.
Two ROs were used, of which one included the wire structure under test.
By comparing oscillation frequencies, the delay of the wire was derived.
Although the authors employed external equipment, they argued that their method could potentially be operated solely on the device under test.
Bruguier \etal \cite{pcfe-bbpm} proposed a non-invasive method to characterize FPGAs performance, analysing the spectra of electromagnetic radiation caused by the ROs.
A similar measurement approach was used by Amouri \etal \cite{aefa-abkb} to explore the impact of elevated temperatures and voltages on the performance degradation of FPGAs.

\paragraph{Discussion}
RO sensors are relatively easy to implement and very versatile, considering that they can provide insights into the effects of BTI and HCI  when combined with methods, such as multi-temperature operational life testing \cite{rpff-bbb-20}.
Additionally, they can be tuned to indirectly measure further quantities than the propagation delay \cite{oshf-zh}.
Although the main sensing principle among approaches is similar, the literature presents multiple approaches to placing the sensors and counting the frequency.
A trade-off exists with RO sensors: longer ring chains cover larger areas and require fewer measurement circuits but decrease sensor resolution.
Care should be taken to avoid overheating the die or stress power rails by using too few stages in the rings, as this results in unrealistic measurements \cite{mimf-pp}.
A downside of ROs is that they measure delays of the FPGA resources that form the ring and not of the synthesized circuits, which may differ depending on the position and routing.
Other techniques measure delays of the existing combinational logic instead, as we describe in the following sections.

\subsection{Shadow Register}\label{subsec:fpga-shadow-reg}

The usage of Shadow Registers (SRs) is a well-studied technique for delay characterization and degradation monitoring.
Sensors are usually placed at the end of critical combinational paths, in parallel to a destination register for detecting late transitions.

\begin{figure}[t]
    \centering
    \includegraphics[width=0.55\textwidth]{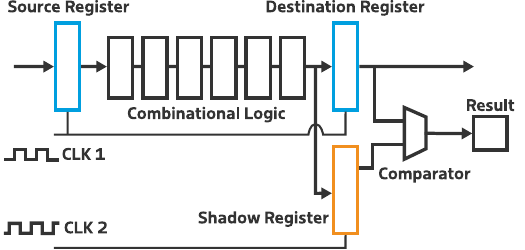}
    \caption{Sensor based on a shadow register negatively skewed to measure the delay of a combinational logic circuit \cite{nsrd-ll}.}
    \label{fig:shadow-reg}
\end{figure}

Li \etal \cite{nsrd-ll}, Valdés \etal \cite{psoc-vfmr}, and Leong \etal \cite{amls-lsts} proposed to place an SR after the CUT that is clocked by a signal skewed from the destination register (see \autoref{fig:shadow-reg}).
By comparing the latched value on both registers and controlling the phase difference of the clock signals, they determine the delay of the CUT.
Li \etal varied the phase difference in runtime to characterize the FPGA propagation delay and built a histogram.
Leong \etal implemented an online concurrent ageing monitoring sensor, which detected when the propagation delay was higher than a predefined threshold.
Valdés \etal included an on/off signal to their concurrent sensor that interrupts the clock, enabling authors to differentiate the type of ageing mostly suffered by the sensor itself: static ageing (continuous monitoring) or dynamic ageing (periodic monitoring).
The sensor functionality was initially tested by operating the circuit under different power supply voltages, which induced a change in its signal propagation delays but did not affect the sensor.
The sensor from Valdés \etal was then validated \cite{dvcs-vfmr} by performing an accelerated ageing process on an FPGA.

The authors reported no significant frequency variations of the clocks, on which the sensor reliability depended after the burn-in process.
Leong \etal tested sensors by increasing the FPGA frequency and reducing the gap time.

Ghaderi \etal \cite{hssd-genb} proposed ageing monitors clocked by a single ``sensor clock``, which set the maximum allowed slack for combinational signals of critical paths.
By injecting the CUT signal and a shifted version of it to an XOR, a positive pulse was generated on each transition of the CUT.
The XOR was latched by a Flip-Flop (FF) and triggered by the sensor clock, thereby detecting invalid transitions whenever the pulse occurred too late.

Pfeifer \etal \cite{drxd-pp} presented an online concurrent delay-fault detection technique for combinatorial circuits.
The authors used the D FFs at the input of on-chip BRAMs as SRs, which map the signals to memory rows for later analysis.
The interconnect introduced a fixed delay between the destination and the SR to control the sensor sensitivity, and an on-board CPU performed the signal comparison.

Wong \etal \cite{sccd-wsc} presented a self-characterization method, with two registers around a combinational CUT, clocked in counter-phase.
An XOR between the CUT output and the SR latched value produced the error signal.
Transitions occurring after the first half of the test clock period were invalid.
The authors leveraged on-chip clock generation to sweep the test clock frequency until the maximum was found.
A non-concurrent circuit for start-up tests was also proposed and optimized in \cite{smcd-wsc}, which stored test results of each region on the FPGA RAM.

\begin{figure}[t]
    \centering
    \includegraphics[width=0.55\textwidth]{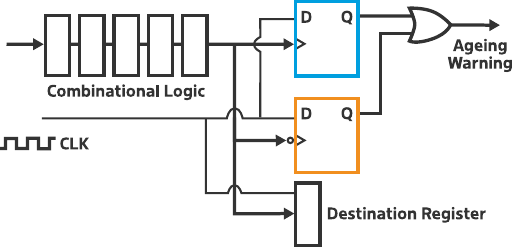}
    \caption{Sensor implemented in \cite{alsa-at} to detect late transitions of combinational logic.}
    \label{fig:ff-sensor}
\end{figure}

Amouri \etal \cite{alsa-at} implemented an ageing sensor to detect late transitions of combinational paths on a Virtex-5 FPGA.
The sensor, illustrated in \autoref{fig:ff-sensor}, was composed of two edge-triggered D FFs clocked by the combinational output, with their inputs connected to the principal clock signal.
Whenever an invalid change in the combinational signal occurred (\ie during the active clock cycle), the sensor output was activated.
Two FFs were used to detect rising and falling signal transitions.
By the addition of independently configurable delay blocks on the combinational output signal and the clock signal, the authors could control the sensor sensitivity (\ie how late after the rising clock signal a change in the combinational output is detected).
When sensitivity is configured to a negative value, the sensor is turned into an early warning monitor, which checks that the signal is stabilized at least by a given time before the clock rises.
In comparison to the previously described work, this method bears the great advantage of not requiring  extra clock resources for the sensor.
On the downside, the approach does not quantitatively measure the propagation delay but rather detects transitions only when slower than a given threshold.

Jiang \etal \cite{afdm-jlth} proposed a similar architecture, but connected the inputs Q of both SRs to a shadow clock signal, which had a phase shift relative to the main clock.
Given that the frequency of the main clock is known, the authors were able to derive the CUT delay by changing the phase angle between clocks and observing the sensor output.

\paragraph{Discussion}
Shadow registers appear more complex to implement and place than ROs, but they provide higher versatility.
These sensors enable measuring propagation delays of application-specific circuits (\eg to characterize chips), as well as implementing late transition detectors.
For the detection of  transitions within a given time window, this method can verify circuit functionality under different conditions and can even run for continuous monitoring of critical combinational paths to detect degradations caused by BTI and HCI.
While ROs either act as probes on unused die areas or temporally replace functioning circuits to test the underlying hardware, SR are able to run in parallel to application-specific combinational logic.
 SR  enable at-speed tests, which is a great advantage as it can seamlessly be added concurrently to applications at the cost of additional resource usage.

\subsection{Circuit Transition Probability}\label{subsec:fpga-transition-probability}

Propagation delay variations can be detected by observing the Transition Probability (TP) of a circuit \cite{tpdm-wdc, dfmm-swsc, damf-swc}.
Consider a combinatorial digital circuit with an output node \(z\).
For each applied input combination, an output value \(z(k)\) is produced.
The \emph{transition probability} of z, denoted \(D(z)\), is the probability of the state changing when the next input stimuli are applied \cite{esac-gdkw} on the following clock cycle.
As \(z\) can only be zero or one,  \(D(z)\) is the probability of \(z\) experiencing a transition between these states:

\begin{align}
    D\left(z\right) = p_z^{1 0} + p_z^{0 1},
\end{align}

\noindent where \(p_z^{0 1}\) and \(p_z^{1 0}\) indicate the probability of \(z\) undergoing the \(0 \to 1\), and \(1 \to 0\) transitions respectively.
From \cite{esac-gdkw}, this probability can be calculated as the relative number of transitions that occurred in an interval of \(N\) clock cycles with \(N \to \infty\).

\filbreak
As an example, \(p_z^{1 0}\) can be defined as

\begin{align}
    p_z^{1 0} = \lim_{N \to \infty} \frac{1}{N} \sum_{k=1}^N z(k) \overline{z(k+1)}.
\end{align}

\begin{figure}[t]
    \centering
    \includegraphics[width=0.55\textwidth]{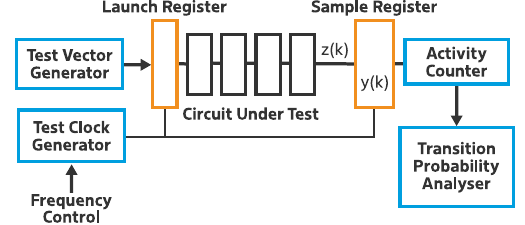}
    \caption{Schematic of the delay measurement method proposed in \cite{tpdm-wdc} based on signal transition probability.}
    \label{fig:transition-probability}
\end{figure}

\noindent If the probabilities are approximated by observing the transitions during a large number \(N\) of clock cycles, we obtain
\begin{align}
	D\left(z\right) \approx & \frac{1}{N} \sum_{k=1}^N \left\{  z(k) \overline{z(k+1)} + \overline{z(k)} z(k+1) \right\}
    \label{eq:trans-prob}
\end{align}

Hence, \(D(z)\) can be estimated by the relative amount of rising and falling edges of \(z\) over a time interval \(N\).

Ghosh \etal derived a theorem \cite{esac-gdkw} which relates the output value probability with the input values probabilities on a combinatorial circuit.
If the input signals have a probability distribution independent of time (\ie they form a stationary process), then the output signal \(z\) will also have this characteristic.
This means for stationary input signals that the TP \(D(z)\) does not change in time.

Wong \etal \cite{tpdm-wdc} estimated the maximum functioning frequency of an arbitrary circuit by measuring its output TP.
With a careful selection of the input signals, they ensured a constant TP of the output under normal operating frequencies.
They performed various measurements at increasing frequencies, up to the point at which changes in the TP could be observed.
The change indicated that the maximum frequency was reached, and the circuit started to fail.
The proposed setup was implemented on a 65 nm Altera Cyclone III FPGA, as shown in \autoref{fig:transition-probability}.
The CUT and registers were clocked from a test clock generator.
On each clock cycle, the test vector generator injected input vectors, which propagated through the CUT generating an output \(z(k)\).
In addition, the sample register captured a sample \(y(k)\) from the CUT at a frequency \(f_{clk}\).
An asynchronous counter recorded the transitions in \(y(k)\) over \(N\) clock periods, later used to estimate \(D(y)\) in the TP analyser circuit.
When \(f_{clk}\) is within the operational range and no faults occur on the circuit, then \(y(k) = z(k)\) and the transition probabilities \(D(y) = D(z)\).
If \(f_{clk}\) is increased above the CUT propagation time, then \(y\) will start to sample values of the previous cycle (\ie \(y(k) = z(k - 1)\)), thus changing \(D(y)\).

The method proposed by Wong \etal was later applied in the study of circuit degradation under accelerated stress conditions by Stott \etal \cite{dfmm-swsc, damf-swc}.
The authors implemented multiple CUTs on a pair of Cyclone III FPGAs, which could be measured using the TP method and allowed to be electrically stressed by an input signal.
Environmental stress with an ageing acceleration factor of \(180 \) was applied to the chips by means of elevated temperature and core voltage, which sped up the NBTI process.
Additionally, the CUT was subjected to electrical stress by controlling its switching activity through the input signals, which triggered NBTI, TDDB, and HCI degradation mechanisms.
Their experiments revealed a circuit speed reduction of up to 15\% by the end of the test schedule.  This stress conditions degraded LUTs stronger than interconnects.
Moreover, the method was verified against a RO-based (see \autoref{subsec:fpga-ro}) frequency measurement.

\paragraph{Discussion}
Measuring changes in the TP of a CUT output allows for detecting its maximum operational frequency.
Unlike ROs (\autoref{subsec:fpga-ro}), this method measures the propagation delay of the FPGA using existing circuits, so the evaluation of the impact of BTI and HCI processes on the application is more direct.
On the one hand, this technique has the advantage of being implementable with common resources and only requires a controllable clock signal.
On the other hand, it requires injecting precise test vectors, which depend on the CUT and need to be stored or consistently generated.
Moreover, as custom inputs are needed, this technique affects the system operation and can only be implemented during a testing period.
 \section{Ageing on Microcontrollers and SoCs}\label{sec:ageing-mcu-soc}

Microcontrollers and SoCs are digital ASICs.
They suffer from similar degradation processes as FPGAs, but their functionality is known from the design process.
This knowledge allows designers to study the expected ageing and the tolerances.
Guard-bands can be placed to counteract chip degradation (\eg limiting operating frequencies) and increase product reliability.

Many applications attempt to reduce costs by utilizing COTS components for longer than the guaranteed lifespans, which may raise critical conditions for highly-dependable use cases.
Despite the wide usage of microcontrollers and SoCs in embedded applications, studies on the impact of hardware ageing on their performance and useful lifetime are missing, in particular for in-field monitoring techniques.
Indeed, most work on measuring degradation builds on utilizing external equipment.
In this section, we present corresponding work, including techniques to assess the impact of ageing on microcontrollers and SoCs.
We focus on COTS devices, which exclude special on-chip ageing sensor.
Our motivation for this is that ASICs including dedicated ageing sensors are not commonly found on the market for embedded devices.

\subsection{Timing Window Violation}

\begin{figure}[t]
    \centering
    \includegraphics[width=0.55\textwidth]{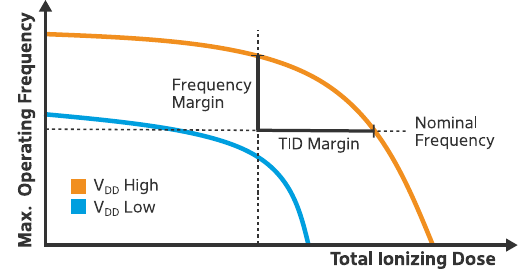}
    \caption{Illustration of the timing window violation hypothesis proposed by Diggins \etal \cite{tidi-dmhk}, in which the maximum operating frequency degrades as the total ionizing dose rises.}
    \label{fig:timing-window}
\end{figure}

Chip ageing can degrade signal propagation times, which also affects microcontrollers.
Many low-end microcontrollers have a dedicated central clock signal and a short execution pipeline.
Under such conditions, a violation of the timing windows, in which data is latched, will cause incorrect data to be propagated down the pipe.

Diggins \etal conjectured a timing window violation hypothesis \cite{tidi-dmhk} (see \autoref{fig:timing-window}) which explains the relationship between \one TID degradation, \two operating frequency, and \three supply voltage.
For each voltage, a maximum operating frequency allows for executing software successfully.
This frequency decreases as the TID increases \cite{stro-sbwm}.
Authors proposed an offline test to measure the impact of TID in the propagation delay of an ATMEGA328P microcontroller, observing violations on timing windows under multiple voltage and frequency conditions while the device was exposed to different values of TID and executing functional tests.
Because the operating core frequency controls the length of the timing windows, software tests were executed at different frequencies.
For each value of TID, the highest frequency at which the test passed was recorded.
The authors reported that by overclocking the microcontroller above its nominal operational frequency, the degradation of the hardware was observable before a failure occurred at the nominal operating point.
This observation bears the potential of developing monitoring techniques that activate predictive maintenance or graceful degradation approaches, such as operating at lower frequencies.

\paragraph{Discussion} The timing window violation hypothesis on microcontrollers has so far only been tested on TID-induced degradation.
However, we envision the necessity to evaluate other ageing mechanisms that affect signal propagation delays.
Although the analysis of timing window violation based on the clock frequency and operating voltage has mostly been tested in  laboratories, great potentials lie in online non-concurrent tests due to the high clock reconfigurability of many modern microcontrollers \cite{rsw-dcrci-22}.

\subsection{SRAM Initial Pattern}\label{sec:soc-sram}

\begin{figure}[t]
    \centering
    \includegraphics[width=0.50\textwidth]{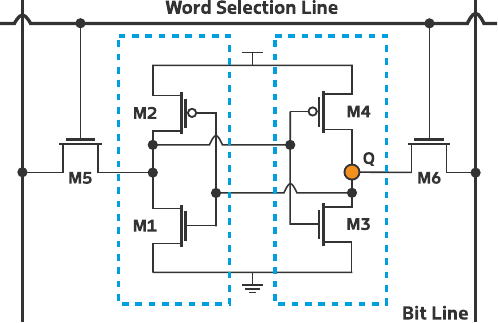}
    \caption{Schematic of a six-transistor static RAM cell circuit. See \cite{rkgs-rlsr-23} for background.}
    \label{fig:6t-sram}
\end{figure}

Many MCUs and FPGAs integrate static RAM (SRAM).
A common implementation of an SRAM cell is the six-transistor circuit (see \autoref{fig:6t-sram}), owing to its relatively low static power usage.
Each cell is composed of two cross-coupled inverters (M1, M2 and M3, M4) and two access transistors (M5 and M6).
The cell has two stable states that represent either a logical 0 or 1, depending on the voltage at \(\boldsymbol{Q}\).
The word selection line grants access to the cell and connects it to the bit line, which transfers data on read and write operations.

When a cell is energized, the initial value is not forced by the bit line; instead, it depends on the mismatch between the threshold voltage of the inverter transistors.
The skew of an SRAM cell is its tendency towards a value when powered \cite{pssf-hbf}.
A non-skewed cell has a small threshold difference, and its initial value is random and depends on noise.
These cells are commonly used as sources of entropy for random number generator systems.
Cells with a moderate mismatch have a tendency towards a particular value, but this can be affected by ageing mechanisms that influence the threshold voltage of transistors, such as BTI.
Finally, fully-skewed cells produce the same initial value with a high probability and can be used to implement SRAM physically unclonable functions.

The Static Noise Margin (SNM) of an SRAM cell is the lowest level of voltage noise that flips its logic state.
As the SNM depends on the transistor threshold voltage, ageing degrades it.
Degraded SNM makes cells more sensitive to electric and thermal noise.
A degradation of \SI{15.02}{\percent} of the SNM has been reported \cite{atac-fn} after ten years of usage, owing to BTI effects.
The skew of a cell is also impacted by BTI and HCI.
Considering a cell that is constantly stressed by retaining a logical 1 on its output, P-MOS M2 transistor in \autoref{fig:6t-sram} is active and undergoes a degradation process that modifies its threshold voltage during this time.
If, upon startup, the gate threshold voltage of M2 is much larger than that of M4, the cell will be skewed towards 0 because M4 will be activated before M2.

The skew change  of SRAM cells over time has been exploited by Guo \etal \cite{zprs-grtf} as a two-phase method to detect recycled SoCs and FPGA chips. This was later refined as a framework \cite{scar-gxrt}.
Authors proposed an initial enrolment phase to detect partially-skewed cells, which would likely change their startup value when aged.
The authors first calculated the probability of starting in 0 and 1 for all SRAM bits, respectively assuming room and high temperature for a given number of startup cycles.
The underlying method foresees that high ambient temperature produces a good prediction for the aged state of the bits.
By calculating the difference between the probabilities and comparing it to a pre-selected gap value, they selected the ``ageing-sensitive`` bits.
A chosen threshold value determined the number of flipped bits required to consider a device as used (\ie aged) during the verification phase.
The gap value used to select the ageing-sensitive bits is not universal, and the authors indicated that it should be determined empirically based on measurements of aged devices beforehand.
Tests on artificially-aged Xilinx Spartan-3 FPGAs showed false accept and reject rates of \si{\num{0}} and \si{\num{0.03}}, respectively, when picking the proper parameters.

Guin \etal \cite{drsa-gwhs} also proposed a method to detect recycled SoCs and FPGAs based on the assumption that SRAM cells are balanced by design.
The authors argued that any bias introduced during manufacturing is random and that the amount of 1s and 0s normally distributes with a centre very close to \si{\qty{50}{\percent}} (\ie within \si{\qty{1}{\percent}}).
Therefore, any significant shift of the mean would be the result of circuit ageing.
The technique requires no initial measurement phase nor a golden model against which to compare.
The approach only requires that the circuit is powered-up multiple times to count the amount of 1s.
The authors suggested that having a large enough SRAM array (\eg \si{\qty{64}{\kilo\bit}}) and taking sufficient measurements (\eg \si{\num{100}}) would minimize the impact of thermal noise during the assessment.
Tests on COTS external SRAM chips showed significant changes in the distribution (up to \si{\qty{14}{\percent}} after ageing) and a slight recovery when stress was removed.

Lanzieri \etal \cite{lkfss-aaesl-23} collected and studied SRAM startup patterns from MCUs of \si{\num{154}} naturally-aged embedded devices deployed on an IoT testbed.
The authors analysed the raw patterns and extracted subtle features, including spatial frequency components, bitwise probability of ones, and bit instability, to unveil correlations with the effective usage times of the devices.
The observation of the extracted features revealed not only linear correlations with the usage time, but also discernible patterns in how the firmware interacts with the SRAM, such as initializing variables to 0 and determining their placement within the memory map.
Finally, the authors trained and compared ML regression and classification models with the result of the feature engineering process, successfully evaluating the application of SRAM startup pattern as a ubiquitous and inexpensive on-chip usage monitor.

\paragraph{Discussion}
Given the ubiquity of SRAMs in various digital integrated circuits, the use of its initial pattern for indicating transistor degradation is an interesting and deployable method that requires no external equipment.
Present work shows that BTI-effects are already detectable on SRAMs after a few hours of accelerated ageing, as well as in long-term naturally aged deployments.
To take further advantage of modifications in SRAM startup patterns, and to integrate them into higher-level health assessment systems, it is required to research correlations between the observed changes and the degradation of functional parameters of interest, such as SNM.

\subsection{Electromagnetic (EM) Characteristics}\label{subsec:soc-em}

Electronic device ageing affects EM characteristics, such as conducted and radiated emissions \cite{emic-bdlb} as well as susceptibility \cite{cmsa-lwhz}.
By subjecting devices to EM Compatibility (EMC) tests, their effective ageing could be determined.

Dawson \etal \cite{ehtm-dfdm} proposed a method to study the effect of high temperature ageing on the EM emissions of a COTS MCU.
Authors proposed to monitor device emissions as a reliable and straightforward measure to monitor ageing, as no electrical contact with the system is required.
An emission test board was built, containing a Microchip PIC MCU with pins connected to loaded tracks passing by RF couplers.
During tests, the software running on the device toggled the pins while a spectrum analyser captured EM emissions and the output voltage was measured at the pins.
The authors reported variations in the emission spectra of artificially aged devices.
Some harmonics of the central toggling frequency increased, particularly the high-frequency harmonics.
It remains unclear whether a shift in the toggling frequency is the culprit.
Additionally, the output voltage on aged devices increased, but there were no conclusions regarding the reasons for the reported behaviour. Although no clear mechanism linking the changes in emissions to device ageing was presented, BTI is a likely candidate due to the temperature-triggered nature of the changes.

EM Robustness (EMR) studies the impact of circuit ageing on EMC \cite{cemc-bbbl}.
Li, Wu \etal \cite{cmsa-lwhz, msaa-wllz} analysed the drift in the EM immunity of an MCU to Electrical Fast Transients (EFT) interference after accelerated ageing.
They aged the device in incremental steps by applying high voltage (150\% above nominal) and high temperature.
The test setup consisted of a probe connecting the device to an EFT signal generator, which injected an interference signal between each ageing step.
To assess whether the device was affected by the signal, errors  were monitored for the executed software.
The authors reported failures, including self-recovering errors, soft errors (reset was needed) and complete damage.
Failure rates increased consistently with the ageing time, as well as immunity degradation (observed as a decrement in the maximum tolerated interference voltage for some pins).
No further analysis was performed relating internal ageing mechanisms under the observed results.
Instead, authors suspected that \one physical parameters of the functional components of the MCU drifted due to ageing, and \two components of the EM interference protection circuitry degraded, thus increasing susceptibility.

\paragraph{Discussion}
Although related work reveals a clear impact of circuit degradation in the EMC and EMR of MCUs, the inner dynamics of this phenomenon are still to be explored.
As MCUs are complex devices, it is challenging to assess the degradation of every subsystem and their impact on the EM spectrum signature.
EM measurements have not been widely developed as ageing monitoring techniques.
This may owe to requiring external equipment and limited understanding of the correlation between system degradation of circuits and the changes in EM characteristics.
This gap opens an interesting research direction for studying correlations between other ageing indicators for MCUs and SOCs  (\eg SRAM startup patterns) and EMC, which could enable indirect measurements of the latter by means of cheaper and ubiquitous sensors that require no external equipment.

\section{Ageing Detection on Power Supplies}\label{sec:ageing-voltage}

\begin{figure}[t]
    \centering
    \includegraphics[width=0.55\textwidth]{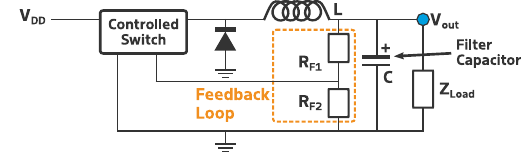}
    \caption{Schematic of a switching-mode power supply in step-down (or buck) configuration, see also \cite{b-pwpsd-12}.}
    \label{fig:buck}
\end{figure}

All embedded systems require supplies to deliver power with stable parameters over the entire lifespan of an application.
Due to their high efficiency (typically more than \si{\qty{90}{\percent}}), Switching Mode Power Supplies (SMPS) are a popular choice to provide a stable DC voltage.
SMPS are commonly composed of a switching transistor, a diode, a component that accumulates energy (usually an inductor) and a capacitor completing an LC filter for a steady DC output.
Depending on their application, they can be built in a step-up (boost) configuration -- reducing current and increasing voltage -- or in a step-down (buck) configuration (see \autoref{fig:buck}) -- increasing the delivered current by decreasing the voltage --.
The controlled switch continuously switches between ON and OFF state, at which ideally is no power dissipation.
The output voltage is regulated by controlling the duty cycle of the switching signal.

\begin{figure}[t]
    \centering
    \includegraphics[width=0.55\textwidth]{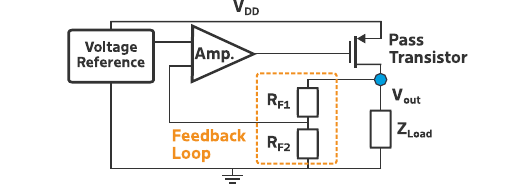}
    \setlength{\belowcaptionskip}{-20pt}
    \caption{Schematic of a generic low drop-out voltage regulator. Refer to \cite{b-ilr-08} for background.}
    \label{fig:ldo}
\end{figure}

Low Drop-Out (LDO) linear voltage regulators (see \autoref{fig:ldo}) can regulate their output even when the input voltage (power supply) is close to the output voltage.
They are commonly found as stand-alone chips in many power delivery networks of embedded systems and integrated into ICs and SoCs.
An LDO is composed of an operational amplifier, a voltage reference, a pass transistor, and a resistive divider which closes a feedback loop.
The amplifier constantly compares the feedback loop voltage with the expected value of the reference.
Based on this value, the amplifier controls the gate voltage of the transistor, which acts as a variable resistance and sets the feedback voltage (and thus the output voltage) to the desired value.
Although simpler than SMPS, LDOs suffer from lower efficiency.
As the voltage regulation is performed by the pass transistor in its ohmic region, it actively dissipates energy.

The most popular choices of power supplies in embedded appliances are built with components that undergo ageing as discussed in \autoref{sec:circuit-ageing}.
In this section, we present several techniques used to detect and monitor the degradation of power supply performances and shifts in their parameters.

\subsection{Power Supply Rejection Ratio}

\begin{figure}[t]
    \centering
    \includegraphics[width=0.45\textwidth]{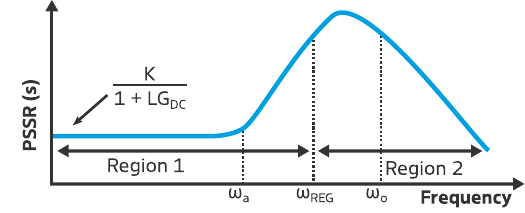}
    \setlength{\belowcaptionskip}{-15pt}
    \caption{Characteristic curve of LDO power supply rejection ratio vs. frequency \cite{rsdl-cgf}.}
    \label{fig:psrr-freq}
\end{figure}

Power Supply Rejection Ratio (PSRR) is an important characteristic of LDO regulators;
it indicates its ability to prevent fluctuations in the output voltage  in the presence of noise from the input voltage (or power supply voltage).
PSRR is the relation between the output voltage ripple \(v_o\) and the input voltage ripple \(v_i\). It is normally measured for the whole operating frequency band of the regulator.
A typical PSRR characteristic is illustrated in \autoref{fig:psrr-freq}; a lower value means a better rejection of the power supply noise.
PSRR in the frequency domain can be modelled as in \autoref{eq:psrr} \cite{rsdl-cgf}.
The constant \(K\) depends on the feedback resistors, the load, and the internal drain-to-source resistance to the small signal of the transistor.
The frequencies \(\omega_a\) and \(\omega_o\) are poles of the PSRR transference equation, and depend on the resistances and capacitances of the transistor, the amplifier, and the load circuit.
\(A_a\) and \(A_o\) are the loop gains of the amplifier and pass transistor, respectively.

\begin{align}
    PSRR(s) = \frac{v_o(s)} {v_i(s)} =
    \frac {K} {\left(1 + \frac{s}{\omega_o}\right) \left(1 + LG(s)\right)}
    \label{eq:psrr} \\
    \intertext{with the feedback loop transfer function,}
    LG(s) = \frac{A_a A_o} {\left(1 + \frac{s}{\omega_o}\right) \left(1 + \frac{s}{\omega_a}\right)} &\quad&\quad
\end{align}

The PSSR curve (\autoref{fig:psrr-freq}) has two regions: Region 1 at low and middle frequencies, and Region 2 at high frequencies.
Both regions meet at the unity bandwidth frequency \(\omega_{REG}\), where the feedback loop gain is one.
Region 1 is governed by the loop gain, which depends on the transconductance of the pass transistor and the amplifier.
As this property is affected by transistor ageing mechanisms (\eg BTI and HCI), PSRR is impacted by circuit ageing.
Region 2 is mainly impacted by parasitic capacitances of the transistor input and the load, while the feedback loop gain plays almost no role.
HCI changes the gate capacitance of transistors \cite{hcic-yyk}, thus modifying PSRR also in this region.

Considering the effect of circuit ageing on PSRR, Chowdhury \etal \cite{aldo-cspm} proposed using this parameter to estimate the ageing of LDOs, as well as ICs, including such components in their power delivery networks, with the goal of determining whether a device has been recycled.
To characterize the impact of ageing in PSRR, the authors designed their own LDO chips in \si{\qty{65}{\nano\meter}} technology.
The LDOs were artificially aged at elevated temperature and power supply voltage.
To analyse the PSRR of the devices and to record the output spectrum, a spectrum analyser and its tracking generator were employed.
After applying AC and DC stress to the LDO, a degradation of PSRR was observed, as well as a shift in the transfer poles.
This shows that both NBTI (DC stress) and HCI (AC stress) mechanisms could be detected by measuring the PSRR degradation.

Based on the clear influence of circuit ageing on PSRR \cite{aldo-cspm}, the authors introduced a technique \cite{rcdl-cgbm,rsdl-cgf} based on ML methods that utilizes this parameter to detect recycled analog and mixed-signal chips and SoCs, respectively.
As a first step, an unsupervised k-Nearest Neighbours (KNN) model was trained and tested with PSRR data from new and artificially aged devices from one vendor.
Next, the authors used a semi-supervised approach, in which they applied the previously-trained models and tested them against other vendors in an attempt to avoid the need of golden data from all vendors.
Both approaches showed promising results in discriminating used and new LDOs, but required golden data and a supervised training phase.
Finally, for unsupervised learning, two cases were tested. \one PSRR data from an unknown device was given to the model together with golden data, and based on the number of clusters returned, the device could be classified as new or aged.
\two The PSRR of the unknown device were measured twice, before and after a synthetic ageing process.
This last approach had the clear downside of being partially destructive and did not show good results.

Acharya \etal \cite{lbor-avf} presented and implemented an odometer based on a modified LDO circuit intended for the ageing-based detection of recycled ICs.
A parallel feedback path to the regulator (\ie a reference path) was added, which remained unused throughout the device lifetime.
Two signals controlled which path is actively used.
A one-label classifier model evaluated the LDO PSRR when using each path to detect degradation in the regulator components.
Besides the external measurements, this method has the drawbacks of requiring a custom regulator design (\ie cannot be applied on COTS components) and employing redundant hardware.

\paragraph{Discussion}
Measuring PSRR changes has proven effective  for detecting ageing of external LDO regulators.
Although Chowdhury \etal claim that the proposed methods~\cite{aldo-cspm, rcdl-cgbm} are applicable to ICs, they were only tested on stand-alone LDOs.
 Measurements of embedded LDOs were solely presented in \cite{rsdl-cgf}, but the applied models could not detect recycled chips.
In addition, the reported results are based on artificially aged devices due to the lack of used chips.
Given its direct relation to transistor parameters, PSRR is a sensible parameter to reveal ageing effects of BTI and HCI on LDOs. Thus far, the literature has only focused on short ageing periods though  -- a few hours under accelerated ageing correspond to a few days of normal operation.
To fully understand the correlation of PSRR with device ageing over longer periods of time, further research is required.
Moreover, methods proposed so far heavily rely on external testing equipment, and no strategy has yet been  presented to deploy testing in the field.

\subsection{Electromagnetic Characteristics}

Ageing not only affects the EM characteristics of microcontrollers and SoCs (\cf \autoref{subsec:soc-em}) but also those of LDOs \cite{esvr-wlsb} and SMPSs \cite{itae-bhb}.
Indeed, as parameters of the voltage regulator shift away from their nominal values due to degradation, so do their EM emissions and susceptibility.

Boyer \etal \cite{itae-bhb} have studied how thermally-triggered ageing mechanisms affect internal passive and switching components of SMPS in step-down or buck configuration, thus modifying its conducted and radiated EM emissions.
To this end, they have artificially aged four samples of NCP3163 devices at elevated temperature.
Although all samples remained operational after ageing, an average increase between \si{\qty{6}{\decibel}} and \si{\qty{20}{\decibel}} in the conducted emissions and \si{\qty{5}{\decibel}} in the radiated emissions spectra were observed over a large frequency range.
The aged inductor was replaced by a new component to determine its impact on the system degradation, which lowered EM emissions according to their initial values.
By means of an impedance \vs frequency characterization of the inductor before and after thermal stress, the authors encountered a reduction in its impedance and quality factor, which was attributed to an increase in the parasitic parallel capacitor caused by a higher core loss as described in \autoref{subsec:ageing-inductors}.

Boyer \etal also performed related experiments \cite{stab-bggd} on synchronous buck converters based on the LT3800 controller.
Measurements of conducted EM emissions were carried out inside a semi-anechoic chamber.
Converter boards and various additional components were aged in an oven at high temperature for two weeks.
The power iron inductor and aluminium capacitor were the most degraded components, followed by the onboard output capacitor and the power transistors.
In addition, the conducted emission measured at the circuit output suffered an increment of up to \si{\qty{15}{\decibel}} between \si{\qty{4}{\mega\hertz}} and \si{\qty{100}{\mega\hertz}}.

As analysed in detail by Wu \etal \cite{esvr-wlsb}, EMI-induced offset is a common failure mode caused by EM noise coupled to the power pin of an LDO.
Due to the nonlinearity of the differential transistor pair at the input of the comparator amplifier (see \autoref{fig:ldo}), noise injected via the power network induces an offset of the output voltage via a rectification effect.
A test bench was built with a signal generator coupled to the voltage source and an oscilloscope monitoring the output and reference voltage.
Various devices were aged for a week by electrically stressing them to evaluate the relation between ageing mechanisms and EM susceptibility.
The devices still functioned, but their EMI robustness decreased.
This indicated that the stress process had indeed triggered the transistor ageing mechanisms in the operational amplifier of the LDO.

\paragraph{Discussion}
Component degradation alters  EM emissions and affects susceptibility of power supplies.
The main culprits are ageing on passive components (via ESR changes on capacitors, or quality factor reduction on inductors), and switching components (mainly due to BTI and HCI).
Although EM characteristics are indicators of these ageing effects, measurements can be troublesome.
EM-based ageing detection methods developed in the literature typically require special setups, procedures, and external equipment, which hinders deployment  as an online monitoring technique.
Moreover, the usage of expensive dedicated instruments contradicts many typical requirements for COTS-based embedded applications, namely low cost, low power usage, and small size.

\subsection{Equivalent Series Resistance (ESR) of Capacitors}

\begin{figure}[t]
    \centering
    \includegraphics[width=\textwidth]{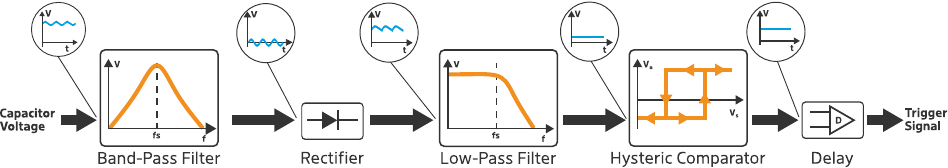}
    3\setlength{\belowcaptionskip}{-10pt}
    \caption{Block diagram of voltage ripple measurement for electrolytic capacitor failure prediction proposed by \cite{ecfp-ccw}.}
    \label{fig:ripple-measurement}
\end{figure}

Capacitors are widely used as low-pass filters to clean noise from power lines (\eg to suppress the switching noise introduced by SMPS as visualized \autoref{fig:buck}).
Capacitor degradation jeopardizes the availability of the circuit.
In fact, electrolytic capacitor failures cause more than 50\% of SMPS breakdowns \cite{fpec-lvgv}.
As digital components require a stable voltage, capacitors are attractive for degradation studies.

The impact of ageing mechanisms on the output voltage ripple of SMPS has been studied in \cite{itae-bhb}.
Several DC-DC switching step-down circuits were subjected to accelerated ageing via increased temperature.
The authors reported an increase between 100\% and 200\% in the output noise after ageing.
It was determined that the main culprits for the decay in the voltage stability were the filtering electrolytic capacitors at the output.
Indeed, the high temperature accelerated the ageing mechanism described in \autoref{subsec:ageing-caps}.
As a consequence, the equivalent serial resistance increased, and the capacitance decayed.

As the ESR usually changes more than the capacitance for a given ageing, the former is preferred for monitoring purposes.
To prevent failures, it is usually a good practice to replace a component whose ESR has increased by 100\% or more \cite{pelc-g}.

Givi \etal \cite{cmdc-gfg}  proposed and implemented an online concurrent monitoring system for DC-DC converters.
Their technique detects ageing at the power supply output capacitor by indirectly measuring its ESR.
The proposed setup required two voltage sensors at the converter: a Rogowski coil sensor on the  inductor and a sensor of the output voltage.
The authors implemented a monitoring system that derived the capacitor ESR on software from measurements of these two voltages using a DS1104 controller board.
By comparing the resistance value with the initial one, the degradation of the component was estimated, and a warning signal could be triggered.

Lahyani \etal argued \cite{fpec-lvgv} that  only the waveform of the output voltage ripple changes when capacitors degrade.
They developed a software-based ESR monitoring method that sampled this ripple and applied a band-pass filter to the signal.
They found that measuring the ripple at the power supply switching frequency is a more realistic method than using an average rectified value, and it reduces the load dependence.

On the other hand, Chen \etal \cite{ecfp-ccw} also measured ESR by observing the output voltage but proposed to filter out the switching frequency.
Following a theoretical analysis, the authors concluded that the amplitude of the output voltage ripple could be accurately modelled as a direct, linear dependence  on the ESR.
Additionally, they showed that the output voltage ripple barely varies when the load current changes because the capacitor is part of an LC filter.
From this, authors derived a method for failure prediction, which measured and processed the capacitor voltage as illustrated in \autoref{fig:ripple-measurement}.
A band-pass filter eliminates the DC component and the power supply switching frequency from the signal.
After injecting the signal into a rectifier and a low-pass filter, a voltage proportional to the capacitor ESR can be obtained.
When compared with a pre-defined threshold voltage, an early warning trigger signal can be generated.
An additional delay circuit serves as mitigation against fake measurements during the circuit startup transient.

\paragraph{Discussion}
Literature agrees that the output capacitor ESR serves as a good indicator to assess the overall health of a power supply, mainly due to the high impact of this component in the correct operation of the system.
Additionally, its measurement is relatively simple.
Given the ubiquity of analog-to-digital converters on most modern microcontrollers of embedded applications, it would be of interest to build in concurrent online test that performed the signal processing completely digitally without extra circuitry.
On the other hand, a drawback of this approach is the high susceptibility of ESR to operating conditions.
Although it has been shown \cite{cmdc-gfg,fpec-lvgv,ecfp-ccw} that changes in the load do not affect the measurement, ESR still changes significantly with temperature and frequency.
Systems need to additionally measure the environment and keep a record of the initial value under different conditions to account for these changes.
 \section{Discussion and Future Directions for Research}\label{sec:discussion}

The body of literature reviewed in this work shows that state-of-the-art research on ageing monitoring predominantly studies degradation effects in a component-wide scope.
In reality, components are not deployed to work in isolation.
Embedded systems are complex and involve various interconnected, interdependent components.
As complex systems, they exhibit composite effects and often present emergent dynamics.
To achieve an understanding of system-wide health, further research is needed.
Questions, such as `How do ageing mechanisms of one component interplay with those of others?' demand an answer.

Following this system perspective, it is equally important  to recognize the impact of low-level physical conditions on the overall system performance.
It is necessary to enhance upper-layer logic with ageing monitoring insights in order to implement more reliable and autonomous systems.
Work on degradation mitigation techniques at upper layers has started \cite{camt-kadk-17}.

The literature represents extensive research towards the implementation and deployment of online tests for FPGAs.
Indeed, their re-programmable nature makes them excellent candidates for synthesizing test resources, which can later be removed during operation.
Ring oscillators are frequently employed as sensors for variability analysis and ageing monitoring, even on more modern FinFET-based devices.
Nevertheless, techniques to assess or monitor degradation on microcontrollers, SoCs, and power supplies have merely focused on offline tests, mainly to understand the impact of ageing processes.
To close this research gap, BIST and SBST for physical degradation effects on these components require further investigation, in particular with a focus on ageing monitors that are deployable online.

We envision that research on dynamically adaptive strategies (\eg dynamic voltage \cite{iruws-kbw-16} and frequency scaling \cite{rsw-dcrci-22}), augmented with degradation information, can help to advance ageing-aware embedded devices without hindering the overall system performance beyond the strictly necessary.
In an attempt to abstract developers from hardware degradation effects, vendors impose design-time guard bands.
Such limits guaranty for embedded components that operate within specifications to account for normal degradation conditions.
Nonetheless, in certain deployment environments degradation may affect components more than expected (\eg enhanced radiation \cite{tidi-dmhk, lbkfs-sdpdf-24}).
Operation outside vendor boundaries can reduce system reliability (\eg higher clock frequencies, lower core voltage, or deployments longer than guaranteed lifetimes), but it can also boost application performance \cite{edfc-kcmp-22}.
Monitoring the device degradation and testing the actual hardware limits may enable such type of operations.
Indeed, by continuously self-assessing the limitations of the components, an ageing-aware dynamic guard band could be determined at which the device still operates safely.

We foresee that adaptive hardware resource management techniques, both at component and system levels, can benefit from degradation monitoring systems.
This would enable better informed decisions on resource allocation, based on task characteristics and underlying hardware conditions.

We have observed a trend towards applying ML  to enhance data produced by performance variability sensing techniques \cite{rcdl-cgbm, iama-lhww-22, lkfss-aaesl-23}, particularly applying traditional models.
In parallel, there is a growing research focus on ML-based predictive maintenance (PdM) \cite{mlpm-csvf-19}.
Still, the field of ML is continuously evolving, and introduces new methods, such as  probabilistic timeseries forecasting to PdM \cite{pfif-asmp-22}.
This includes models which can analyse live online data, perform zero-shot inferences without historical data, as well as online re-training.
We consider that research on these ML models in combination with built-in online ageing monitoring systems has strong potential to further enhance the reliability and automation of embedded system deployments with PdM.

In this context, tinyML is under active research as a low-power edge alternative to cloud-based ML solutions \cite{cstml-ytbb-23}.
This technology is already being applied in the domains of anomaly detection and PdM, but mostly for sensor data \cite{nomla-ms-21}.
We envision that tinyML combined with on-device degradation monitoring can enable fully ageing-aware autonomous systems, by allowing devices to collect and self-asses the results of built-in tests without requiring third parties nor upstream connectivity.

A research program to address the identified gaps in the field could be guided along the following questions.
\begin{enumerate}
\item Is it possible to reuse the working principles of offline tests on SoCs, microcontrollers, etc.~to derive in-situ online tests of physical device degradation?
\item Can we identify lightweight, easily measurable indicators that faithfully represent the  overall health of a system?
\item Can we derive reliable system functions for health monitoring and self-assessment from these early indicators?
\item Is there a versatile lifetime model that predicts system failures based on the indicators available from the self-assessment?
\item How to efficiently design a predictive maintenance architecture that can leverage low-level insights from built-in embedded sensing and ML inference based on historical data?
\end{enumerate}
 \section{Conclusions}\label{sec:conclusion}

In this work, we reviewed the dominant degradation mechanisms that affect the basic components of embedded systems, which include a particular perspective on the trend of transistor miniaturization.
We systematized methods for detecting hardware ageing of the elementary  building blocks used in COTS embedded systems: FPGAs, SoCs, microcontrollers, and power supplies.
This review was motivated by the increasing deployment of such embedded systems in varying contexts of criticality, as well as in harsh operational environments.
Many deployments are hard to access, either because they are widely distributed in the IoT or because the environmental conditions prevent accessibility (\eg highly radiated tunnels of particle accelerators).
Systems in these environments should operate autonomously and should also be able to monitor and diagnose themselves.

Degradation detection techniques are also useful to assure component quality and identify counterfeit hardware. Recent chip shortages have increased quality concerns and raised the need for acceptance testing.
Counterfeit is not limited to parts manufactured by different entities; it is estimated that \si{\qty{80}{\percent}} of counterfeits are recycled \cite{cicd-gdt-14} and potentially under-performant chips \cite{arfd-asi, zprs-grtf, aldo-cspm}.

With this overview, we envision to foster future research and development towards self-aware embedded systems, which not only can detect the degradation of individual hardware components, but also are able to assess their overall health status and predict a remaining lifetime.

\balance

\bibliographystyle{ACM-Reference-Format}

\end{document}